\documentclass[journal,12pt,onecolumn,draftclsnofoot,]{IEEEtran}

\usepackage{graphicx}
\usepackage{amssymb}
\usepackage{booktabs}
\usepackage{amsmath}
\usepackage{enumerate}

\ifCLASSOPTIONcompsoc
  \else
 \fi

\ifCLASSINFOpdf
 \else
  \fi

\begin{document}

\title{Dependable Structural Health Monitoring Using Wireless Sensor Networks}
%Enabling 
%Energy-Efficient and Fault-Tolerant Health Monitoring of Civil Infrastructures Using Wireless Sensor Networks

%Energy-Efficient and Fault-Tolerant Structural Health Monitoring in Wireless Sensor Networks
%Enabling Distributed Fault-Tolerant WSN Framework for Structural Health Monitoring
%Distributed Fault-Tolerant WSN Framework for Structural Health Monitoring
%On Distributed Fault-Tolerant Detection in Wireless Sensor Networks
%Automatic
 %Fault Tolerance in Wireless Sensor Networks for High Quality Structural Health Monitoring
%High Quality Structural Health Monitoring With Fault-Tolerant Wireless Sensor Networks

%\author{Md Zakirul Alam Bhuiyan,~\IEEEmembership{Member,~IEEE,}
        %Guojun Wang,~\IEEEmembership{Member,~IEEE,}
        %Jiannong Cao,~\IEEEmembership{Senior~Member,~IEEE,} % <-this % stops a space
				%and  Jie Wu,~\IEEEmembership{Fellow,~IEEE} % <-this % stops a space

\author{Md Zakirul Alam Bhuiyan, ~\IEEEmembership{Member,~IEEE,}
        Guojun Wang, ~\IEEEmembership{Member,~IEEE,}
				Jie Wu, ~\IEEEmembership{Fellow,~IEEE,} and
				Jiannong Cao, ~\IEEEmembership{Fellow,~IEEE,}

\IEEEcompsocitemizethanks{
\IEEEcompsocthanksitem M. Z. A. Bhuiyan is with the School of Information Science and Engineering, Central South University, Changsha, China, 410083, and the Department of Computer and Information Sciences, Temple University, Philadelphia, PA 19122. E-mail: zakirulalam@gmail.com.\protect
\IEEEcompsocthanksitem  G. Wang is with the School of Information Science and Engineering, Central South University, Changsha, China, 410083. E-mail: csgjwang@gmail.com (Corresponding Author). 
\IEEEcompsocthanksitem J. Wu is with the Department of Computer and Information Sciences, Temple University, Philadelphia, PA 19122. E-mail: jiewu@temple.edu.
\IEEEcompsocthanksitem J. Cao and X. Liu are with the Department of Computing, The Hong Kong Polytechnic University, %Kowloon, 
Hong Kong. E-mail: \{csjcao,csxfliu\}@comp.polyu.edu.hk.

%A preliminary version of this paper appeared in IEEE SRDS'12 [4].
} % <-this % stops a space
\thanks{}}

% The paper headers
%\markboth{IEEE Transactions on Dependable and Secure Computing, ~Vol.~XX, No.~XX, XX~XXXX}%
%{M.Z.A Bhuiyan \MakeLowercase{\textit{et al.}}: Dependable Structural Health Monitoring using Wireless Sensor Networks}

\IEEEcompsoctitleabstractindextext{%
\begin{abstract}
As an alternative to current wired-based networks, wireless sensor networks (WSNs) are becoming an increasingly compelling platform for engineering structural health monitoring (SHM) due to relatively low-cost, easy installation, and so forth. However, there is still an unaddressed challenge: the application-specific dependability in terms of sensor fault detection and tolerance. The dependability is also affected by a reduction on the quality of monitoring when mitigating WSN constrains  (e.g., limited energy, narrow bandwidth).
 We address these by designing a dependable distributed WSN framework for SHM (called  \texttt{DependSHM}) and then examining its ability to cope with sensor faults and constraints. 
We find evidence that faulty sensors can corrupt results of a health event (e.g., damage) in a structural system without being detected. 
More specifically, we bring attention to an undiscovered yet interesting fact, i.e., the real measured signals introduced by one or more faulty sensors may cause an \emph{undamaged location to be identified as damaged (false positive)} or a \emph{damaged location as undamaged (false negative)} diagnosis.
 This can be caused by faults in sensor bonding, precision degradation, amplification gain, bias, drift, noise, and so forth. In  \texttt{DependSHM}, we present a distributed automated algorithm to detect such types of faults, and we offer an online signal reconstruction algorithm to recover from the wrong diagnosis. Through comprehensive simulations and a WSN prototype system implementation, we evaluate the effectiveness of  \texttt{DependSHM}. 

\end{abstract}

}

\maketitle

\IEEEdisplaynotcompsoctitleabstractindextext
\IEEEpeerreviewmaketitle

%\vspace{-1cm}
\section{Introduction}  
Wireless sensor networks (WSNs) consist of a number of sensor  nodes that can collaborate with each other to perform monitoring tasks. WSNs have been widely deployed  on  the ground,  vehicles,  structures, and the like for enabling various applications, 
e.g., target detection, scientific observation, 
safety-related, and traffic monitoring \cite{04,102,3040,103,130,132,133,3041}. A WSN typically consists of a large number of resource-limited sensor nodes working in a self-organizing and distributed manner.  Sensor nodes Applications of WSNs include military sensing, wildlife tracking, traffic surveillance, health care, environment monitoring. 
Recent work has explored that WSNs can be a compelling platform for engineering structural health monitoring (SHM), due to relatively low-cost, easy installation, and so forth \cite{106,530,522,507,101}.
%Structural health monitoring (SHM) is treated as a potential application of WSN in recent years, due to the significant benefits of low-cost, easy to deploy and maintain since no cables are required,  short setup time and low maintenance cost.
%
 In a typical SHM system, the interest is in monitoring possible changes (e.g., damage, crack, corrosion) on physical structures (e.g., aerospace vehicles, buildings, bridges, nuclear plants, etc.)  and providing an ``alert'' at an early stage to reduce safety-risk. This prevails throughout the aerospace, civil, structural, or mechanical (ACSM) engineering communities. 

Both ACSM and computer science (CS) communities have already addressed numerous challenges/requirements, including data acquisition, compression, aggregation, damage detection, distributed computing.
%Both ACSM and CS communities have already made sufficient contributions in this multidisciplinary application, considering numerous challenges/requirements, including data acquisition, compression, aggregation, damage detection, distributed computing, energy consumption, etc.  %However, the WSN for SHM system could be more effective if requirements, such as distributed computing, energy cost, sensor fault detection and tolerance, are incorporated. 
However, there is  still an unaddressed challenge: \textit{the application-specific dependability}, which is the ability of a WSN providing  application-specific meaningful monitoring results under sensor faults.
%
%Network survivability is the ability of a network keeping connected under failures and attacks, which is a fundamental issue to the design and performance evaluation of wireless ad hoc networks.
%
 %the application-specific dependability in terms of sensor fault tolerance. 
%If we want to put our confidence on a WSN-based SHM system, it is important to have the system be dependable. 
Particularly, such a system should be able to detect the sensor data faults online and take recovery actions immediately to avoid meaningless monitoring operations% or monitoring degradation
. In fact, dependability is highly desired in a WSN-based SHM, as an ``alert'' about a structural event conveys a serious concern with public safety and economic losses. 

%In fact, dependability is highly desired here as we requires an immediate action when there is an ``alert'', which is a serious concern with public safety and economic losses.
% This  is because data may be  corrupted  further during the wireless transmission to the BS that can bring inaccuracy in identifying the fault detection and  offline monitoring.

On the one hand, SHM algorithms in wired sensor networks used by ACSM are generally centralized/global-based \cite{522,2008,2007,501}, in which they may not need to seriously consider data collection quality and synchronization errors, etc. This is because they may not often handle data losses or mismatch, as there are no issues like poor wireless connectivity, narrow bandwidth, and energy constraints. The dependability is affected by a reduction on the quality of data when mitigating the constraints.
%Moreover, we often find that network assumptions are made for SHM systems due to lack of consideration for CS requirements, such as energy, connectivity, and bandwidth constraints in WSNs.
%WSN constraints, e.g., energy cost, fault tolerance, strong connectivity, etc%wireless bandwidth
 %In order 
%The ACSM communities do not concern about CS requirements, such as energy, connectivity, bandwidth, data delivery, etc, in WSNs.
% and have no much concern about energy cost. %,  a typical example is Ting Kau Bridge
%Those algorithms work on the raw measured data, which is no longer a single value (e.g., ``0/1'') but a sequence of data with a size of more than hundreds of KB. Thus, they incur huge  communication overhead to the traffic-sensitive WSNs. 
Moreover, once data from the WSN is collected at a centralized base station (BS), it becomes complex to scrutinize all the collected data (including faulty signals). 
%In the CS community, it is expected that the 	resource optimization of WSNs must be tightly correlated with the respective applications, e.g., SHM. 
%
% that SHM technique require. %, especially when a lot of sensors are deployed or the structure is large. 
%In contrast, it is commonly believed that for the resource-limited WSNs, system design would be more efficient if the application requirements are incorporated. 
%
%Before deploying a WSN on a real structure, an  energy-efficient distributed framework is essential to address the constraints in the WSN. 
%
%To make good use of WSNs for SHM, a purely distributed and energy-efficient framework for WSN is essential.   

%Focusing on the second issue, in order to make meaningful conclusions in SHM, the quality of monitoring health status must be ensured under the sensor faults. 
%
%The problem is that when a set of sensors are placed at strategic locations for %long-term 
%monitoring under harsh environment, many sensors are likely to exhibit unreliable behavior. The biggest problem induced by sensor faults is that the signals from those sensors can be erroneously interpreted as a consequence of a structural damage. Such false interpretations lead to false decisions like unnecessary additional inspections %of the structure 
%associated with down time of the structure and related costs. %Unlike wired sensor networks based SHM where energy and fault tolerance are not big factors, WSNs are energy-constrained, their batteries cannot usually be recharged or replaced, and they are prone to failure.
%
%As an important research issue, 
On the other hand, significant efforts have been made for specific fault types in WSNs \cite{3001,3035}. 
%Many of the them may not be directly applicable to SHM systems. 
Some prominent schemes, namely, \emph{decision fusion (or 0/1 decision)}, \emph{threshold-based} decision, %\emph{value fusion},
 \emph{heartbeat reception} have been suggested for fault-tolerant phenomenon (such as an event) detection problems \cite{3001,3022,3011,3014,3015}. %%3018
These often use simplified data and few measurements to adequately detect certain faults. However, they are not able to function properly in an SHM system, since SHM algorithms use totally different methods to detect a damage event. For example, the algorithms need raw measured signals rather than the decision fusion, and the analysis of signals (vibration, strain, damping, etc.) that requires a substantial knowledge from ACSM domains (e.g., finite element model updating, Eigen matrix, mode shape properties) \cite{1040,101,530,1033}. 
We have evidence from experimental settings that when there is a change in structural health properties (as shown Fig. 1a), 0/1 decision schemes tell sensor 5 is faulty, but they cannot tell what happen (faulty signals or damage event) around sensors 4 and 6. %% [3014]
%Thus, removing faulty signals from measured signals and identifying what happens exactly in a structure is a challenging task.
%Furthermore, in an investigation on SHM system dependability, we have found that those 0/1 and threshold-based decisions making schemes are not showing satisfiable performance, as shown in Fig. 1.  
Regarding all these issues  above, a question might be posed: is it possible to have a dependable SHM system using WSNs?

The answer is positive. In this paper, we design a dependable and distributed WSN framework for SHM (called  \texttt{DependSHM}) that jointly considers ACSM and CS requirements. In  \texttt{DependSHM}, we propose an algorithm to detect sensor faults efficiently under the constraints of the WSN. 
%Application specific, especially, SHM specific fault-tolerance under WSNs is still a overlooked research problem. 
Dependability in WSNs suffers from various types of faults, including, transceiver failure,  link errors, security attacks (e.g., collusion), etc \cite{3035,3036}. Numerous efforts are being published every day in handling these  fault types. 
Instead, we are interested in some types of sensor faults that are common but difficult to identify: \emph{sensor debonding (when a sensor partially or completely debonds from the host structure), faulty signals, faults in offset, bias, precision degradation, and the amplification gain factor of signals, noise faults, node missing or failure}.

 %: faulty sensor readings.
%\begin{itemize}
	%\item Sensor debonding---it is a very common fault in SHM that occurs when a sensor partially or completely debonds from the host structure;
	%\item Faulty signals---these are caused by precision degradation and breakage, especially in capturing vibration signals  from the structure; 
	%\item There are faults in offset, bias, precision degradation, and the amplification gain factor of signals;
	%\item Noise faults are due to noisy readings;
	%%Noise faults that are caused by longer duration noisy readings;
	%\item There is also node missing or failure.
%\end{itemize}
%% compromized node, security fault
 
%Sensors with these faults seem to work properly, to communicate to neighbors, to exchange \emph{heartbeats}, but they return incorrect values or decisions. We think that most of the sensor data faults fall within these four faults models.
%
%Most of the sensor data faults fall within these fault models. Since these types of faults cannot be easily identified, they directly interrupt a WSN system from detecting damage. 
Most of the sensor data faults fall within these fault models and they directly interrupt a WSN system from detecting damage.
Sensors with some of these faults seem to work properly, to communicate to neighbors, to exchange \emph{heartbeats}, but they return incorrect readings or decisions. Under any of the fault occurrences in a practical SHM, we discover a fact that goes to SHM system dependability: \textit{both faulty and non-faulty sensors can generate abnormal signals or decisions} (i.e., remarkable changes in the measured signals). 
%
%This indicates that there is a possibility of both structural damage/crack and sensor faults occurring at the same time.
 The difficult part is that sensor data, the only available information, will be affected by both structural damage and sensor faults. 
We further discover an interesting fact that such a  possibility can cause  an \emph{undamaged location to be identified as damaged (false positive)} or a \emph{damaged location can be given undamaged (false negative)} diagnosis. When we transform these false positive and negative rates into a structural health event detection ability as the performance of system dependability (as shown in Fig. 1b), we find that those decision based and current SHM schemes do not perform well. 
%%As a result, distinguishing the damage in the presence of sensor fault is a challenging task. 
%As a result, removing faulty signals from the measured signals and identifying what happens exactly in a structure is a challenging task.

%During monitoring, it is impossible to guarantee that whether there is ``damage'' or ``no damage'' if there is a ``faulty sensor''. 
%On the one hand, readings from faulty sensors are geographically independent. On the other hand, readings from sensors in the close proximity are spatially correlated [19][3].  %In this paper, we propose a general approach for detecting arbitrary types of faults. The scheme includes 
We use a new general measurement, \emph{mutual information independence} (MII),  between two signals $u$ and  $v$ from two different sensors for evaluating results in the absence of the ground truth. We think that mutual statistical information can be used as an \emph{indicator} to decide on a sensor fault detection in conjunction with damage detection. We attempt to reconstruct faulty sensor signals using Kalman filter techniques so that if there is damage, it can be recovered after the reconstruction. This does not require any costly %cost-ineffective 
actions, including sensor grouping, faulty sensor avoiding, masking, isolating, or replacing. %). %Besides SHM, the original idea of the work also applied to other WSN application, such as environmental monitoring, intrusion detection, etc. 
%This recovery is applicable to any kind of faulty signals caused by numerous sensor faults.

%%%%%%  Figure  %%%%%%
\begin{figure}[tb]
\begin{center}
\includegraphics[scale=0.6]{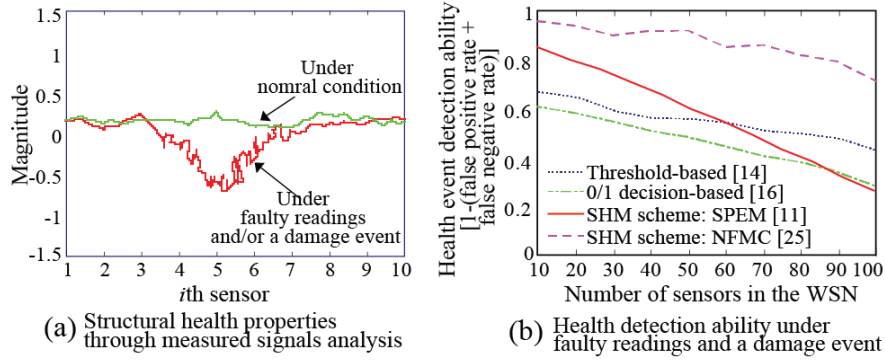}
\end{center}
\vspace {-4mm}
\caption{Investigation of the dependability performance of different schemes in structural health monitoring (SHM).}
\vspace {-4mm}
%\vspace {2.5cm}
\end{figure}
%%%%%%%  Figure  %%%%%%

Our major contributions in this paper are as follows:  
\begin{itemize}
\item We study a WSN-based SHM system dependability problem and %, namely damage detection under sensor faults. 
design  \texttt{DependSHM} to address the problem. This task is by no means easy, as it requires multi-domain knowledge and is associated with optimizing WSN resource constraints.%, including wireless communication, network traffic.  .  
% since distributed computing is essential due to the limitation associated with wireless communication and network traffic. 
%
%In literature, very  few such solutions exist to date.  
%where health status can be computed in a distributed manner, 
%Very  few solutions exist  to date  regarding  meeting  real-time requirements  in  WSN. 
%To date such solutions are limited for SHM.
%
\item 
%To the best of our knowledge, we are the first to investigate a light-weight  method of an automatic sensor fault detection. 
We propose a non-faulty data collection algorithm, by which we utilize an %automated 
online faulty sensor detection algorithm based on the function of MII. 
% sensor mutual information independence (MII). 
Although we focus on sensor faulty signals in  \texttt{DependSHM}, MII does not rely on a particular fault type. %Thus, it can be easily applied to other WSN applications.
\item  %To tolerate the faults 
In  \texttt{DependSHM}, we present a  recovery algorithm to reconstruct faulty sensor signals based on the Kalman filter technique. The recovery is directly applicable to any kind of spatially and temporally correlated signals that are caused by numerous sensor faults in a WSN-based SHM system.
%. %, meaning that it does not rely on a particular fault type 
\item We evaluate  \texttt{DependSHM} via simulations using real data sets, adopted from a %an existing 
SHM system deployed on the GNTVT structure \cite{1030}.
%We achieve almost the same monitoring quality under sensor faults with that of the civil engineering, while reducing significant energy cost in the network. 
We implement a prototype system developed by the TinyOS \cite{1514} running on the Imote2, and verify it on a test structure. %Experiments conducted using physical structures demonstrate that our system has the ability to accurately detect damage in the presence of sensor fault. 
The results show that a careful use of recovery from faulty signals in \texttt{DependSHM} is effective and can lead to a %fully-automated 
dependable WSN-based SHM system. % for data collection and reconstruction of real world and non-stationary signals in WSNs. 
%This work hints the advantages of a pure scheme to \emph{a cyber-physical system} (CPS) that closely integrates the design of computing systems and physical structural engineering. 
%The results validate that the use of recovery from faulty signals is effective and can lead to a fully automated fault tolerant monitoring system. 
\end{itemize}

%\footnotetext[1]  { http://www.cse.polyu.edu.hk/benchmark/}

This paper is organized as follows. 
Section 2 reviews related Work.
Section 3 provides system models and problem formulation. 
Section 4 presents the \texttt{DependSHM} framework.
The faulty sensor detection algorithm is in Section 5. 
Faulty sensor signal reconstruction is detailed in Section 6. 
 Performance evaluation % and experiment are 
is outlined in Section 7. 
%We conclude this paper in Section 8.
Section 8 concludes this paper.{

%\footnotetext[2] { Appendices are attached to the supplementary file of this paper.}

%%%%%%%%%%%%%%%%%%%%%
%%%%%%%%%%%%%%%%%%%%%

\section{Related Work} 
%\subsection{Dependability in WSN-based SHM}

\textbf{Dependability in WSN-based SHM.  } WSNs have been widely suggested and validated in experimentation for SHM system by  both the ACSM and CS communities in recent years \cite{530,522,101,2008,2007,1033,106,108}. Existing schemes already have sufficient contributions to ACSM and CS requirements \cite{102,106,530,522,507,101,2008,2007,501,1033,2010,109,111,3039,3039}, but they suffer from the dependability problem. 

On the one hand, generally data can be corrupted  at four stages, namely acquisition, processing and local decisions, wireless transmission, and the final analysis at the BS. 
%Any of the stages can bring inaccuracy in  monitoring, leaving the monitoring system undependable. %Research attention is nowadays turning towards drawing a meaningful scientific inference or a quality of monitoring from the high-quality of collected data in any of the stages.
 Among them, the most important stage is the acquisition stage that can ensure the quality of sensor readings in WSNs at the beginning.
 %Plenty of WSN deployments for generic applications (will be described later) have already observed faulty sensor readings caused by incorrect hardware design, improper calibration, security attack, or by low battery levels \cite{3035,3009,3010}. Given these observations, it is  impossible to have a perfectly calibrated WSN in practice.
The quality is also affected when application-specific requirements are considered, including high-resolution data, raw data, non-faulty data, dependable and real-time decision-making to analyze actual structural health conditions. These additional requirements are traditionally guaranteed by using wired networks.
% Traditional wired network almost satisfy these requirements, as there is no big concern about sensor faults, energy constraint, bandwidth constraint, etc. 
To make WSNs effective alternatives to wired network system instruments, a first step in this direction is SHM system dependability in terms of detection of faulty sensor readings and a collection of non-faulty readings at the BS, and then dependable monitoring results. In this paper, we take such a step. 

%\subsection{Work from Generic WSN Applications Related to SHM Dependability}
\textbf{Work from Generic WSN Applications Related to SHM Dependability.  }   Fault tolerance in WSNs has been studied extensively by researchers in computer science \cite{3001,3022,3011,101,523,3011,3015}.  The application background is largely event/target detection in battlefield surveillance, environment monitoring, etc. %%3018
The general objective is that when some sensor nodes give faulty readings, how to achieve the correct detection of an event/target over a specific region. 
Among them, a large part of the schemes on fault detection are off-line and centralized-based.
% We discuss several pieces of work that are closely related to our scheme.
Most schemes rely on various detection methods, including correlation analysis, 0/1 decision, value fusion,  decision rules or threshold. Some more details can be found in our earlier work \cite{3011,106,523,3015}.

Although dependability support by fault detection and tolerance problem in SHM looks similar to the problem of making binary  ``0/1'' detection decisions or value fusions \cite{106,3011,3014}, it is fundamentally different from them. 
To check the validity of this assumption and the dependability, we have conducted WSN-based SHM experiments on a physical structure (the settings are described in the later part). 
For these experiments, %the prototype SHM system is installed on the structure, and sensor data obtained during both normal (i.e. non-faulty signal)  and abnormal (faulty signal) system operation are used to train the dependability capabilities of the SHM system. Thereupon, 
faulty signals are injected into sensor nodes to validate the system’s capabilities to autonomously detect and tolerate the fault. 
As shown in Fig. 1, these schemes do not show a satisfactory performance in SHM; specifically, they cannot identify what exactly occurs in a WSN-based SHM. We can see that when there is a change in structural health properties (as shown Fig. 1a, a 0/1 decision scheme \cite{3014} tells sensor 5 is faulty, but it cannot tell what happens (faulty signals or damage event) around sensors 4 and 6. There are also changes in signals of sensors 5 and 6. Those existing schemes show here a high rate of \emph{fault positive} and \emph{false negative} rates, resulting in a low system dependability. When we consider WSN-based system dependability as the structural health detection ability, we can see that these schemes show low dependability performance, as shown in Fig. 1b. The methods of detecting faulty sensors by measured signals, removing faulty signals from the measured signals, and then  identifying what happens exactly in a structure are different from the methods in those schemes.

 Numerous techniques towards the area of fault detection and isolation (FDI) have been proposed, e.g., model-based techniques, knowledge-based techniques, or a combination of both \cite{3029}. There are also techniques on fault-tolerant data aggregation that deal with faulty sensor readings caused by security attacks, such as node compromising, collusion \cite{3035,3036}. They use some filtering algorithms (e.g., iterative filtering) at the upper-stream nodes (e.g., cluster head) to remove the faulty signals. Though the algorithms seem to be applicable for our case, we could not justify them. However, it may be difficult to apply such filtering algorithms at a upper-stream node once such a  high-resolution big acceleration data from a number of nodes reaches  at a upper-stream node and the upper-stream node filters all the raw data. Various constraints in WSNs and application requirements might be an issue in them.

%\subsection{Work from SHM Applications Related to SHM Dependability}
%
\textbf{Work from SHM Applications Related to SHM Dependability.   }
On the contrary, there are also fault detection schemes from ACSM engineering domains \cite{2014,2011,2015,3031}. 
The concepts in most of them are associated with system failure detection dating back to the 1980s.
 %The most popular scheme is to make use of analytical redundancies in the system. 
Here, the failure does not imply to faults in a WSN system, but to faults (e.g., damage) in a physical systems. 
FDI concepts (described previously) have also been implemented in a number of engineering disciplines, such as ACSM, to improve the availability and reliability of SHM systems \cite{3031}. 
 However, these are centralized and computationally-intensive, and usually developed in wired networks. 
 %these are proposed for wired sensor networks, which are centralized, computationally-intensive, and hence are not applicable to resource-limited WSNs. 
%The most important part is that these schemes also mainly handle the structural faults (e.g., damage) and the sensor fault detection is the minor part.
 %Moreover, algorithms similar to these assume that the structure itself is healthy and is not able to disambiguate structural damage from faulty readings. 

A noteworthy WSN deployment method for SHM applications called SPEM \cite{501}. SPEM  nicely explains the ACSM and CS requirements and is verified on the real structure. It adjusts the quality of sensor locations to better fit WSN requirements; meanwhile, the adjustment  satisfy ACSM location-quality requirements. We have verified SPEM under sensor faults in simulations and found that the SHM dependability performance in SPEM drastically decreases from 87\% to 28\% as the number of sensors in the WSN increases, as shown in Fig. 1b. This is just because of a lack of the dependability support.

To the best of our knowledge, as the first step, we have worked towards the WSN-based SHM application-specific dependability, and have got preliminary results  \cite{106,523}; and this paper is an extension. Our first work is about sensor fault detection algorithm and structural damage event in WSN-based SHM \cite{523} that works on Natural Frequency extraction and Matching in Clusters (NFMC for short), and then tackles faulty sensor readings.  However, it has several shortcomings, described in \cite{106}. 
%Detection task is carried out by using knowledge of natural frequency of vibration. Frequency sets, which include only the pick frequencies (referring to Fig. C2) collected from different sensors in a cluster, are matched. A frequency set matching algorithm is provided. However, the same pick frequency sets are not possible to achieve since the real measurement can vary from structure to structure, sensor to sensor, and time to time. Unmatched frequencies are assumed to capture as a consequence of sensor's faulty signals and is deleted from the frequency set. The normal assumption is that unmatched frequencies are as a result of faulty sensor signals and is then deleted from the frequency set. We think that the probability of the actual damage detection is somehow affected by such an assumption (for more details, see Appendix C).
%%
After improving the shortcomings, we check the dependability performance by NFMC, as shown in Fig. 1b. Its dependability performance falls between 96\% and 76\%, which is much better than all other schemes. However, such a performance is still not enough to put our confidence in a WSN-based SHM system.  
 
%Besides, the faulty sensors are isolated from the network in NFMC. Both transmit all sets of natural frequenqcies and sensor isolating/masking requires significant energy cost, which are not verified in NFMC. These shortcomings need to be further improved. Later, some of the shortcomings have been improved in another scheme called FTED (on a fault-tolerant event detection) \cite{1069}. This includes i) distributed extraction of features for faulty node detection, ii) iterative faulty node detection, and ii) distributed event detection. 

As an extension, this paper includes several aspects. 
(i) We deal with the \textit{problem of SHM system dependability} and design \texttt{DependSHM} for the problem. (ii) We propose a WSN framework to observe the dependability by  removing faulty sensor data from structural damage data and by finding a fault \emph{indicator} based on MII%,  which is sensitive to sensor fault but is insensitive to damage, and then implement damage detection algorithm
. To ensure the dependability, we devise a new method for tolerance to a missing or failed sensor.
We then address the challenge of when both \emph{structural damage} and \emph{sensor fault} occur at the same time, and identify what exactly happens in a structure.  Generally, it is not easy to make sure that there is a \emph{faulty sensor} but there is \emph{no damage}.
 %These are based on structural mode information (shape) computation (refer to Appendix B$^2$) rather than natural frequency considering WSN constraints.
 %Because mode shape is more sensitive to damage detection and is computationally-efficient, it includes fewer amounts of noise and data transmission than that of natural frequency. 
Particularly, we attempt to make \texttt{DependSHM} efficient for recovery from the sensor faults that occur for a short duration; thus, masking or isolating the sensor (as required in NFMC) is not needed. 
(iii) For the faulty sensor, we present a process flow of the Kalman filter and a graphical representation of it for the signal reconstruction. 
(iv) The motivation for the dependability, clarifications on system models, and SHM-specific terms are given. 
(v) A detailed performance evaluation and new results are presented.

%%%%%%%%%%%%%%%%%%%%%%%%%%%%%%
%%%%%%%%%%%%%%%%%%%%%%%%%%%%%%

%%%%%%%%%%Section%%%%%%%%%%%
%%%%%%%%%%Section%%%%%%%%%%% System Model
\section{Models and Problem Formulation} 
In this section, we  provide some necessary models and definitions. Then, we formally formulate our problem.

\subsection{Network Model} 
We assume that a set $P$ of $m$ wireless sensors is in charge of performing different types of application tasks (e.g., sensing the vibration, strain, and damping signals, pressure, temperature, etc., in the context of SHM) and sending its measurements to neighboring nodes. 
 A reference 2D building model is shown in Fig. 2a, where sensors (white circle) are deployed on it and a remote monitoring center or  BS station location (colored circle) is at a remote place. Fig. 2b shows the traditional WSN framework for SHM (which is similar to the framework in \cite{501}), while Fig. 2c shows the proposed distributed, dependable WSN-based SHM framework, \texttt{DependSHM}.  

\begin{figure}[tb]
\begin{center}
\includegraphics[scale=0.36]{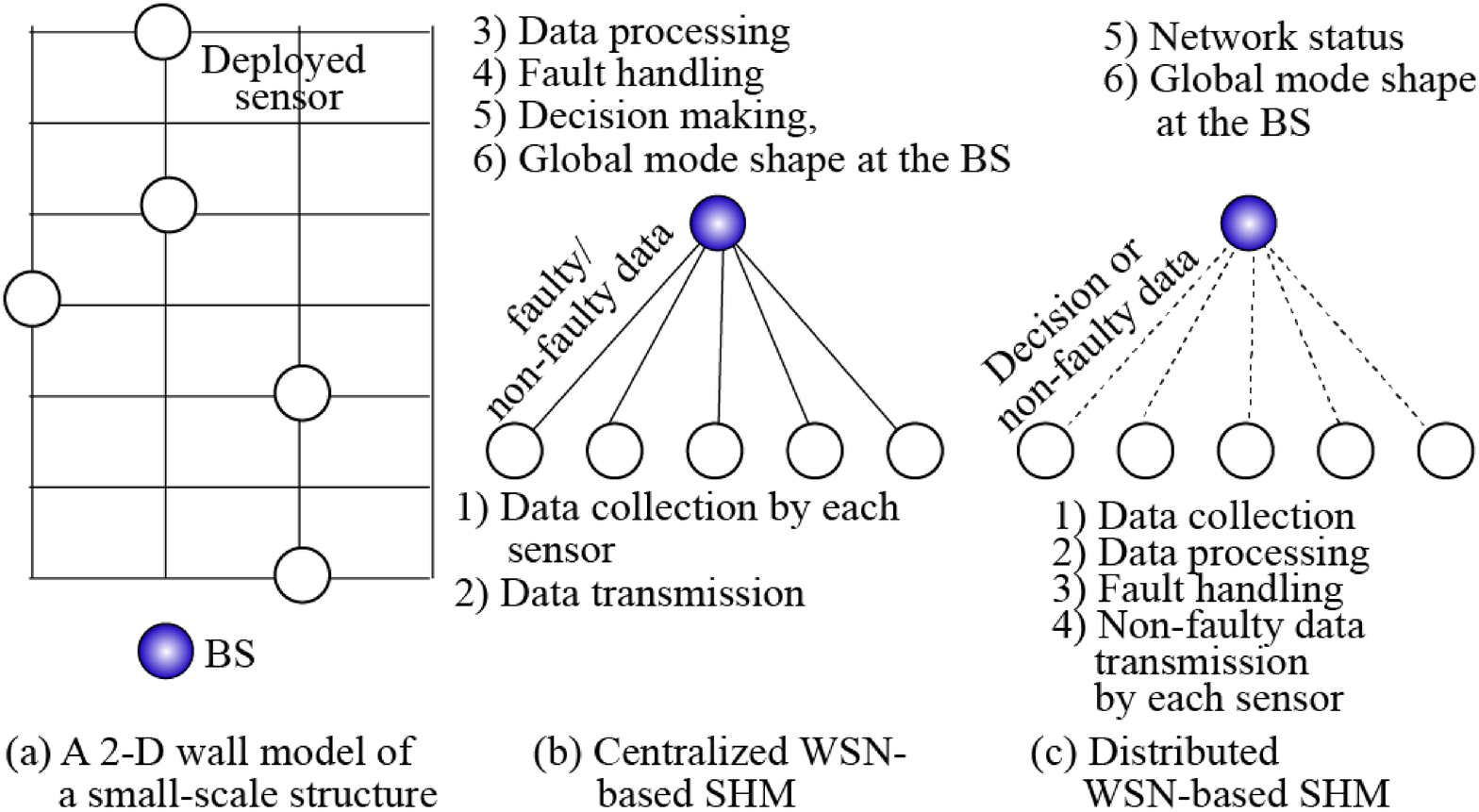}
\end{center}
\vspace {-3mm}
\caption {WSN-based SHM frameworks.} %\label{fig2}
\vspace {-4mm}
%\vspace {4cm}
\end{figure}

%\vspace {1mm}
%\textbf{Definition 1 }[Natural Frequency]. \emph{Every structure has a tendency to vibrate with a larger amplitude at some frequencies than others, assuming there is no outside interference. Each such frequency is  a natural frequency, denoted by $f$}.
%
%\textbf{Definition 2} [Structural Mode shape]. 
 %\emph{Upon excitation, the vibrations are a combination of several harmonics (or at a specific frequency of vibrations), known as modes. Each mode deforms the structure into a particular spatio-temporal pattern known as a mode shape, denoted by $\Phi$}.

Consider that the set of sensors is deployed on a structure
%For simplicity, we assume a simple and yet general enough model that is widely used in the community. 
%We assume that $m + 1$ wireless sensor nodes are deployed 
by finding locations from a set of candidate locations of the structure; $L = \{l_0, l_1, l_2,\cdots,l_m\}$, where sensor $i$ is placed at location $l_i$, and $l_0$ is a suitable location of the BS. For high-quality monitoring results, we follow engineering-driven sensor deployment method \cite{101,501,104,105}.
 %This deployment may be performed by a traditional uniform/random method. However, according to the ACSM engineering domains, to achieve a high-quality of monitoring in WSN-based SHM, engineering-driven sensor deployment should be followed \cite{101,501}.
Regarding \texttt{DependSHM} in Fig. 2c, sensor $i$ can be allowed to acquire data, analyze it locally (prepare natural frequency if needed),  identify faulty readings%and make decision on sensor faults
, and finally compute mode shapes locally or transmit the non-faulty raw data %directly 
to the BS (see Appendix B for the natural frequency and mode shape definitions). 
%(see Definitions B1 and B2 in Appendix B for natural frequency and mode shape). 

%Let $R$ be the sensor communication range.
 Let %$R$ be the communication range, 
$R_{max}$ and $R_{min}$ be the maximum and minimum communication ranges of a sensor, respectively. $R_{min}$ is used to maintain local topology, where a number of sensors is allowed to share their signals with their neighbors % in a neighborhood denoted by $D(u)$
 for damage detection, also used for fault detection. The intention of adopting adjustable communication range is to reduce energy cost for transmission. Note that Imote2 sensor platform supports discrete power levels \cite{1516}. Two local topologies can be seen in Fig. 3; each sensor can be overlapped by one or more sensors. When a sensor communicates to the BS directly, $R_{max}$ is used. Each sensor corresponds to a vertex in a network graph denoted by $G$, and two vertices are connected in $G$ if their corresponding sensors communicate directly. The graph $G$ is called the communication graph of this WSN.
 %We assume that a connection are ``reliable'': when a node $v_i$ sends some data to a neighboring node $v_j$, the total message cost is only 1. 

%A sensor's measured readings is faulty if it is significantly independent of its neighbors. Sensors with faulty readings are called faulty sensors. Let $y_i$ denote the reading of the sensor $i$. Instead of a ``0/1'' binary decision [12] [9], $y_i$  is assumed to represent the actual reading factor that is measured by (7)% or other methods proposed in literature [8]
%. 

\subsection{Sensor Faults}  
\subsubsection{Fault Model} 
%We focus on a set of sensor faults that occur in a real WSN-based SHM system that corresponds to faulty sensor readings.
We focus on the following set of sensor faults that occur in a real WSN-based SHM system:
\begin{itemize}
	\item Sensor debonding---it is a very common fault in a WSN-based SHM %system 
	that occurs when a wireless sensor slightly or completely debonds/detaches from the host structure. This affect is seen in terms of  accurate vibration capturing from the structural response. 
	\item Faulty signals---these are caused by precision degradation, breakage, etc., especially in vibration signal capturing.
	%from the structure. 
	For example, a sensor reports a constant value for a large number of successive samples, where the constant value is either very high or very low compared to the ``normal'' or ``reference''  value.
	\item There are faults in offset, bias, and the amplification gain factor of signals. For example, the offset fault is due to calibration errors in sensor signals, which differs from the normal value by a constant amount, but the sensor readings still exhibit normal patterns.
	\item Noise faults are caused by longer duration noisy readings that affect a number of successive samples.
	\item There is also node missing or failure.
\end{itemize}
%% compromized node, security fault
 
Sensors with these %types  of 
faults seem to work properly (except for the last type), to communicate to neighbors and exchange messages, but they return incorrect values or decisions. 
%These types of faults except the last type are common in WSNs but difficult to detect. 
 We %intend to 
tackle these faults in \texttt{DependSHM}.
% this work.

\subsubsection{Fault Detection Model}  
We assume that sensor $i$ exchanges its signals with its neighbors in each sampling instant $t$ in $T_d$, where $T_d$ is the monitoring round, i.e., the time is divided into discrete sampling periods. In each period, $i$ broadcasts its current readings to one-hop neighboring nodes using $R_{min}$. Besides the readings, $i$ may be enabled to make a decision locally or recover the mode shape and forwarded it to the BS. 
%
%By following (7), let $y_i^t$ denote the actual reading of a sensor $i$ at each sampling period,  $t \in {T}$, $t = 1,2,\ldots$, where $T_d$ is a monitoring period. % and $t$ is each round of monitoring. %For simplicity, we express $y_i^t$ as $y$.
The signal $y_i^t$  measured by a faulty sensor at $t$ is subject to %the measurement 
noise effect $\sigma$. Then, 
%the measured output of $i$ is given by:
let $y$ be the measured output reading that would be transmitted to the neighboring nodes, which is given as follows:
\begin{equation}  
 y = y_i^t  + \sigma % 
\end{equation}

The measurement noise, $\sigma $, for non-faulty sensors may be a small random noise in practice. However, it also can greatly affect damage event detection.
Consider that a subset $N$ of sensors is non-faulty at time $t$% in the WSN
. In SHM, when all of the sensors are non-faulty, it is easily possible to estimate the  mode shape from the signals. However, if a sensor is faulty, it is possible to produce predicted signals for the mode shape by using neighbors' signals, correlation statistics, and the extent of $\sigma $. 
Suppose that damage may occur at %any 
time $t$, anywhere in the structure. 
%Consider the health status is \emph{damage} somewhere in the structure. 
A subset $D \subset P$ of sensors around the damage area is possibly able to detect the damage. Some of the sensors from $D$ may provide faulty signals. %i.e., some sensor may be faulty in $D$.
 Thus, to detect %the 
faulty sensors, the sensors in $D$ split into two further subsets: 
\begin{center} 
\emph{N}= sensors assumed to be non-faulty\\
\hspace{-7mm} \emph{F} = sensors assumed to be faulty
\end{center} 
Note that these two sets are disjoint so that   \\
\begin{center}  
  $N \cap F = \{ \} $ and $N \cup F = D$, where $D\subset P$ % $N \cup D = P$ 
  \end{center}
	 
Generally, sensors anywhere in the WSN can be faulty/failed. However, %considering the structural monitoring environments, 
we put an emphasis on continuous monitoring and on those sensors whose signal %behaviors 
have changed significantly (due to a fault/damage% occurrence
). %We further assume that only those sensors that are around a damage area may fail during an extended period of operation.
 %Nonetheless, there may be multiple damage event occurrences in the structure.

%\vspace{1mm}
{\textbf{Definition 1} [MII: Mutual Information Independence]. \emph{A  function denoted by $\omega ()$ is defined by the quantify of how much the measurement correlation between %the 
sensor nodes in $N$ and %the 
sensor nodes in $F$ deviate from the correlation model.}
%\vspace{1mm}

We state the MII function as an indirect vibration signal measurement. Assume that a prior correlation model $C$ of  $x_P^t$ presents \cite{1080}. $C$ can be given as a reference set by all immediately-stored data in the sensor local memory after WSN system initialization. %In our experiment, Imote2 sensor provides 32MB space, although we do not utilizes such large space. 
The MII function between two signals of sensors $i$ and $j$ at time $t$ in $D$, is given as follows. % $(y_i^t, y_j^t, C)$.
 
\begin{equation}
\omega (y_i, y_j, C) 
\end{equation}
Consider that the sensors in $N$ and $F$ capture vibration signals and broadcast their measured signal sets $y_{N}$ and  $y_{F}$, respectively. %Hence, MII function is given by: % 
 Thus, MII %function 
between the two sets of signals of $N$ and $F$ is given as follows: %%%[1324]
\begin{equation}% 
\omega (y_{N}, y_{F}, C) 
\end{equation}
Given $R$ consecutive signals, the MII function estimates the correlation between $y_N$ and  $y_F$ at time $t$ as:% 
\begin{equation}% 
\Delta (N,F) = \sum\limits_{t = 1}^R {\omega (y_N} ,y_N,C) - \sum\limits_{t = 1}^R {\omega (y_F} ,y_N,C)
\end{equation} 
 $\Delta (N, F)$ is achieved by reducing the deviation between non-faulty sensors in $N$,  and by maximizing the deviation between non-faulty sensors in $N$ and faulty sensors in $F$. $D$ can be controlled by the system user considering $R_{min}$, neighborhood size, or network density. 
 %Intuitively, non-faulty sensor readings should be consistent with each other, whereas the readings of faulty sensors should be inconsistent. % with non-faulty readings.
 %Note that for generality we do not assume that the faulty sensorsˇ readings are uncorrelated. The false alarm rate (false negative) and the detection rate (true negative) are used as the two performance metrics to evaluate the decision of $N$.
% 

\begin{figure}[tb]
\begin{center}
\includegraphics[scale=0.35]{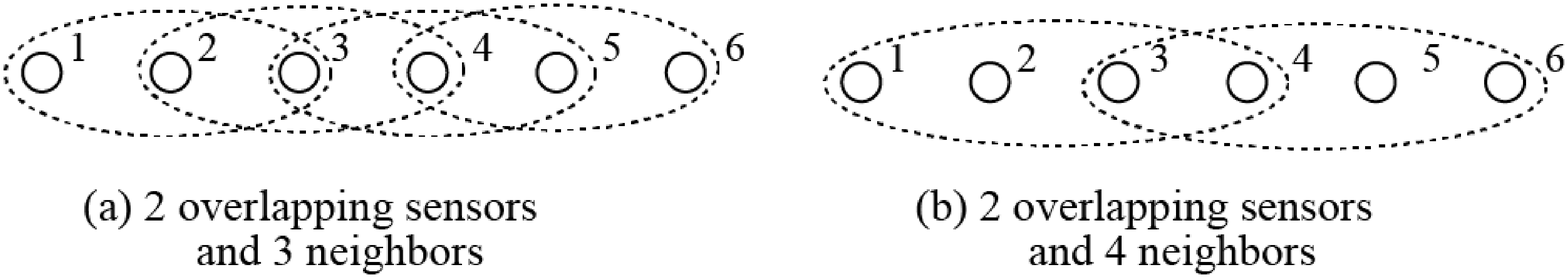}
\end{center}
\vspace {-3mm}
\caption{Topologies with different numbers of overlapping sensors in different neighborhoods.} %\label{fig2}
\vspace {-4mm}
%\vspace {2cm}
\end{figure}

\subsection{Energy Cost Model (cost($e_i$))}  
One important objective is to minimize the energy cost of the network. % regarding the sensor fault detection and recovery. 
 Let $cost(e_i)$ denote the total energy cost of sensor $i$,  including measurement, computation, transmission, and overhead.  
Consider a shortest path routing model \cite{101,501}; there is a path from sensor $i$ to neighboring sensor node or the BS $j$: $q=z_0,z_1 \ldots z_k$. 
Sensor $i$ propagates the data to them. We can find the $i$th hop sensor on each path and calculate the amount of traffic that passes along on the paths within each round of monitoring data collection ($T_d$, $d$=1,2,\ldots, $n$). Then, the $cost(e_i)$ can be decomposed into the following four parts:
\begin{equation} 
{\mathop{\rm cost}\nolimits} ({e_i}) = {e_{T}} + {e_{comp}} + e_{samp}+e_{oh}   
\end{equation}

Here, i) $e_T$ is the maximum energy cost %per bit 
for data transmission over a link between a transmitter and a receiver. % We use a standard energy cost model for packet transmission \cite{1525}. 
ii) $e_{comp}$ is the energy cost for  processing data  locally, e.g., computing equation (6). iii) $e_{samp}$ is the energy required for a sampling cost of $M$ data points \cite{523,513}. iv) $e_{oh}$ is  the additional overhead for some causes, e.g., fault detection, signal reconstruction, copying  data to a local buffer, and %other possible sources (e.g., 
network latency. 
%See Appendix C for the remaining part of explanation of this measurement model and localization error adjustment.
See Appendix A for an extended version of the energy cost model.

\subsection{Problem Statement} 
\hspace{2mm} \textbf{Given}: 
a set $P$ of $m$ sensors %and bandwidth, 
and a BS, which are deployed over a physical structure and are involved in monitoring structural health that is reported to the BS.
%\item An event, fault alarm rate $F_{\mathbb{C}} (h,T_{d} )$, and  $T_d$

%%
%%%a network of 
%%a set $P$ of $m$ sensors with limited energy %and bandwidth, 
%%and an external BS, which are deployed at strategic locations over a physical structure and are involved in monitoring structural health that is reported to the BS.
%%%\item An event, fault alarm rate $F_{\mathbb{C}} (h,T_{d} )$, and  $T_d$

\textbf{Find}:
a subset $D$ ($\subset P$) of sensors that involves in detecting structural damage event such that the sensors in $N (\subseteq D)$ are non-faulty and the sensors in $F = D-N$ are faulty,
%  A subset $R$ of $P$ such that the sensors in $R$ are non-faulty and the sensors in set $F = P-R$ are faulty.
subjected to the following constraints: 
\begin{itemize}
	\item Data delivery: $\forall q = {z_0}\ldots z_k$ used for data delivery, where $\overline {q[j-1]q[j]} \le R_{max}$, $j=1\ldots k$; 
	\item  Connectivity: $\forall i=1\ldots m$,  $\exists q  = {z_0}\ldots z_k$, $q[0]=l_i$, $q[k]=l_0$; 
	\item Structural modeling.
\end{itemize}
  
\textbf{Objectives}:
 minimize  $\Delta (N, F)$ and minimize $cost(e_i)$, and maximize dependable mode shape computation.

%Given:
%\begin{itemize}
%\item A network of a set $P$ of $m$ sensors with limited energy and an external BS, which are deployed at strategic locations over a physical structure and involved in monitoring structural health and report it to the BS.
%%\item An event, fault alarm rate $F_{\mathbb{C}} (h,T_{d} )$, and  $T_d$
%\end{itemize}
%
%Find:
%\begin{itemize}
%\item
%A subset $N (\subseteq D)$ of sensors which detect a structural damage such that the sensors in $N$ are non-faulty and the sensors in subset $F = D-N$ (where $D \subset P$) are faulty.
%%  A subset $R$ of $P$ such that the sensors in $R$ are non-faulty and the sensors in set $F = P-R$ are faulty.
%\end{itemize}
%
%Objectives:
%\begin{itemize}
%\item Minimize  $\Delta (N, F)$.
%\item Minimize $e_T$
%
 %%maximize $T$
%\end{itemize}

%%%%%%%%%%%%%%%%%%%%
%%%%%%%%%%%%%%%%%%%%
%\section{Dependable WSN-based SHM Framework: \texttt{DependSHM}}% 
\section{\texttt{DependSHM} Framework} 
In this section, we present  \texttt{DependSHM}, the dependable WSN-based SHM framework. 
%The whole framework 
It is divided into three stages: distributed framework for structural health event detection (e.g., damage, crack, corrosion), faulty sensor detection, and faulty reading reconstruction. 
%We first start with the event detection process.

\subsection{Basics of Structural Event Detection Algorithm} 
The central focus of SHM is the detection and localization of events (considering damage) within various types of structures. Generally speaking, SHM techniques %for detecting and localizing damage 
rely on measuring structural responses to ambient vibrations or forced excitation. 
Ambient vibrations can be caused by earthquakes, wind, passing vehicles, or forced vibrations, or can be delivered by hydraulic or piezoelectric shakers. It can also caused by a damage occurrence.

A variety of sensors, e.g., accelerometers, strain gauges, or displacement 
can be used to measure structural responses. SHM techniques infer the existence and location of damage by detecting differences in local or global structural responses before/after a damage occurs. The responses %of large structures 
are usually comprised of frequencies in the tens of Hz, and can be sensed using relatively inexpensive low-noise MEMS-based accelerometers. %Structures can be excited by ambient vibrations caused, for example, by wind or passing heavy vehicles
%.
 Increasingly, the ACSM communities are becoming interested in active sensing techniques \cite{530}, which measure structural responses to forced excitations.
 In order to identify the damage, two necessary structural characteristics are important:  mode shape ($\Phi$) and natural frequency ($f$).  %

%\textbf{Definition 1 }[Natural Frequency]. Every structure has a tendency to vibrate with much larger amplitude at some frequencies than others. Each such frequency is called a natural frequency denoted by $f$. $f$ is internal vibration signal characteristic of structure and is different for different structures (such as from building to bridge, from indoor to outdoor).
%
%\textbf{Definition 2} [Mode shape]. 
%When subjected to external forces, the response of a structure is conceptually similar to the response of a vibrating string or structural components such as a metal plate. Upon excitation, the vibrations is a combination of several harmonics (or at a specific frequency of vibrations), known as modes. Each mode deforms the structure into a particular spatio-temporal pattern known as a mode shape denoted by $\Phi$.

\subsection{%Structural Response and 
Mode Shape Computation at Each Sensor} 
Each type of structure (aircraft, building, bridge, etc.) has a tendency to vibrate with much larger amplitude at some frequencies than others. 
%A specific vibration pattern at a specific frequency is called mode shape. 
$f$ and $\Phi$ rely on structural material properties, geometry, and assembly of its constituent members. %Traditionally, mode shape is analyzed for structural damage detection. 
%
%\textbf{Definition 3 }[Damage] A damage in some of members of a structure results in a change in mode shapes and frequencies induced in the structure.
%
We use \emph{state space model}, which is widely accepted by ACSM communities for capturing structural dynamics to compute $\Phi$ \cite{1040,530}. We mention  the process we use for $\Phi$ computation, which is the same process is used for designing faulty signal reconstruction (after fault detection). The state space matrices for a finite-dimensional linear structural dynamic system can be succinctly obtained by the linear differential equation: %[[1816]]

\begin{equation} 
M\ddot x + Kx + \sigma = F(x,t) % 
\end{equation}

Here,  function $F(x,t) $ is the response of the structure over a period of interest at certain sensor locations, where $x$ is the structural response at time instant $t$. $M$ and $K$ are the matrices %% for C for damping
of  mass and stiffness coefficients of the various elements of the structure, respectively$^3$. $\sigma$ is the signal to noise.  
%%% see [CS papre [1559]
%Damping is neglected in the state space model. In any case, faults in the sensors will only be identified if they cause changes in the response of a greater magnitude than the errors in the estimated mode shape. Undamped mode shapes are usually estimated accurately from the model.  In any case, the modes with low damping, having approximately real modes, will be excited most strongly.
%
%\footnotetext[3] {Damping is neglected in the model (6) considering individual measurement estimation.  In any case, (a) faults in the sensors will only be identified if they cause changes in the response of a greater magnitude than the errors in the estimated mode shape, and (b) the modes with low damping, having approximately real modes, will be  strongly excited. Thus, undamped mode shapes can be  accurately estimated by (6). 
%}
%
In (6), damping is neglected for an advantage in detection  (see Appendix C for details).

In traditional SHM algorithms%, [1559]
, the state space model is computed in a centralized/global fashion. We argue that it could be quite costly for the resource-limited WSN. 
Considering SHM as a big data application, to make use of the WSN for SHM%, flexibility, and versatility 
,  we mitigate this problem by allowing each sensor to work only with local structural responses rather than the global. For this purpose, 
We modify the model,  %Thus, an extent of damage detection can be at the local sensors. 
 considering the implementation of the model for each sensor location. 
 %Each sensor computes the state space model locally. 
To reduce the system order, a transformation of the state space into mode coordinates is necessary. This transformation is derived by determining a diagonal matrix, which contains a certain number of Eigen frequencies covering the natural frequency ranges of interest. By applying the mode transformation, based on the mass normalization  $\{\phi_i\}$, the refined $\Phi$ is given by:  %%% [see 1816]
\begin{equation} 
x = \Phi h  
\end{equation}
where $h$ is the mode participation factor. We have, % 
\begin{equation}% 
\ddot h_i + {\delta _i}{h_i} = {H_i}
\end{equation}
(7) implies the refined response of the structure that is a sum of the responses in each mode. We can express it more explicitly %  
 as follows:
\begin {equation} 
x = \sum\limits_{i = 1}^n {{h_i}{\phi _i}} {\rm{ }}
\end{equation}
where $\delta _i$  is the $i$th eigenvalue, and  ${H_i} = \phi _i^Tf$  is the $i$th mode of responses under force or ambient excitation input. 
The summation is given over all of the $n$ modes of the structure that are measured by the individual sensor. Typically, only the lower modes are important because the force excitation is concentrated in these modes.  %%[see 1536]
Each sensor extracts its local ${\Phi}$ that can be used for both faulty signals and damage event detection. The method of extraction, including the difference between $f$ and ${\Phi}$ can be found in Appendix B.

\emph{Is it Possible to Compute $\Phi$ under Sensor Faults:} In practice, $\Phi$ is greatly affected by a faulty sensor signals (see Appendix B for more detail), % or by node missing/failure, 
especially when a sensor is placed at a optimal location \cite{101,501}. If a signal is detected as faulty, the measured signal is reconstructed directly for the actual mode shape% computation
, by collecting the signals from the sensor location or reference signals. We use neighboring sensor nodes' signals for detecting a faulty sensor and reconstruct its signal. In this work, not all the sensors' signals using (6) will be measured. 
 The measured output of a sensor $i$ at time $t$, $y_i^t$, can be obtained by:  
\begin{equation}  
y_i^t = \textbf{Q}x 
\end{equation}
where $\textbf{Q}$ is the measurement matrix. 

\begin{figure}[tb]
\begin{center}
\includegraphics[scale=0.3]{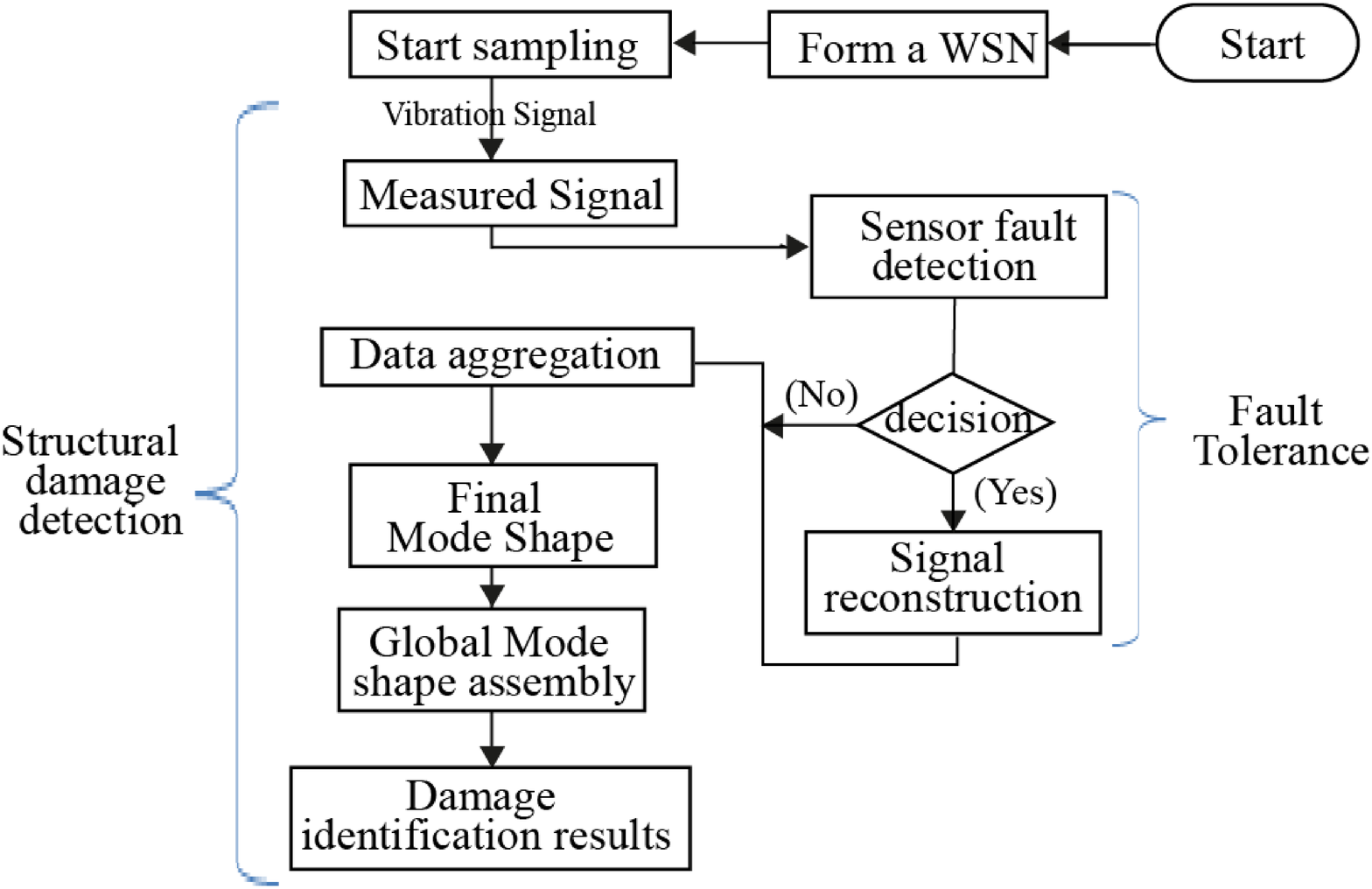}
\end{center}
\vspace {-2mm}
\caption{Providing the dependability in the WSN Framework: % showing
 the step by step process of sensor fault tolerance and damage detection.} 
\vspace {-3mm}
%\vspace {5.8cm}
\end{figure}

\subsection{General Overview of \texttt{DependSHM}} 
%In order to provide fault tolerance in the WSN and take into account the specific aspects of the application, we design the WSN-framework and application-specific fault tolerance. 
Fig. 4 summarizes the whole WSN framework for SHM. Once the WSN starts operating, in each monitoring round $(T_d)$, a number of signals is measured at each sensor. Based on the measured signals, each sensor identifies faulty sensors by using MII. If any faulty sensor is detected, the neighboring sensors reconstruct the signals, while a faulty sensor itself can do the same task if it still works. It is highly possible that a sensor exhibits faulty behavior temporarily in many cases. %, such as hardware/software problem, path-loss problem and so on.
 However, if a sensor fails or it is missing from its location, we still guarantee the signal reconstruction for the sensor. We do not assume to isolate or deploy a new sensor as it is a costly task doing so. 

% in this work. Practically, it is a costly task to isolate or replace a sensor. 

In addition, for the high-quality SHM, we must provide monitoring information of each sensor location, since ACSM communities mostly deploy sensors at optimal locations \cite{101,501}. Each sensor locally computes the final $\Phi$ and transmits to the BS directly% (see Fig. 5)
, which is a relatively small amount of data. The BS assembles all the received final $\Phi$s and identifies the structural damage. There is a high possibility that the BS does not receive any of $\Phi$ caused by data packet-loss or others. If sensor forwarding fails, the BS still has the final results received through the neighboring sensor nodes. In addition, we allow each sensor to keep the final results %(or mode shape)
in the local memory until a sensor receives an acknowledgment from the BS. Each set of raw $\Phi$ is of a number of KBs while each final result or $\Phi$ computed by a sensor is a number of bytes. %The Imote2 sensor is with plenty of space (32MB).
%Transmitting all sets of frequencies to the BS require tens of KB data
 In this framework, each final $\Phi$ received from the neighbors is not processed.

%% 
%%%================
%%%+===============

\section{Faulty Sensor Detection} 
This section describes non-faulty sensor data collection and faulty sensor detection algorithms,  according to the model described earlier. 
%We first provide the faulty sensor reading indication, then we discuss the algorithms. %the some faults that this algorithm can not detect.

\subsection{Data Collection and Faulty Sensor Indication} 
We assume that sensors are likely to generate abnormal signals. The signals are measured by the vibration, which may be incorrect compared to the neighbors, previous signals, or reference signals.
We first show data collection at every sensor in the WSN. A subset of sensors,  say $D$ of sensors that are in a sensor's $R_{min}$%n a neighborhood $D(u)$
, share their data with each other, and participate in  faulty sensor detection. 

Algorithm 1 simply presents the data collection method based on the neighborhood% $D(u)$
. %The constructed $D(u)$ has the maximum node degree at most a constant. 
%A sensor propagate their readings within $D(u)$ %the %a multi-hop neighborhood
 %of the sensor. 
While theoretically this procedure involves multi-hop communication, consider the fact that for SHM application, the radio communication range of current sensor nodes exceeds the area in which sensors gather signals. We limit sensors to communicate within the one-hop neighboring nodes. We think that multi-hop communication is not mature enough. % $d$. %, all nodes in $D(u)$ can be scheduled to transmit once in constant $\beta  = \Theta (d)$ %%[1320]. time-slots without causing interferences to other nodes in N. We take $\beta$  time-slots as one round. 
 The nodes that are one-hop away from the BS will directly send the data; otherwise, the data is sent through one or more intermediate nodes. 
%We also note that here we omit describing  the $D(u)$ construction and node degree for brevity.\
 In Step1 of Algorithm 1, every sensor acquires signals captured from vibration responses of the structure, and buffer them temporarily. Then, it transmits and receives the measured signals. The sensors check if there are any faulty sensors, i.e., a sensor with faulty signals via Step2. 

Step3 executes Algorithm 2. 
When a remarkable change appears in a sensor's signals, there is a possibility that a sensor is faulty. 
%The mutual information independence (MII) 
The MII is used to detect faults. % by providing decisions for sensor fault status besides damage detection. 
Let us consider the statistical dependency between the two sensors' signals quantified by MII. $\omega$ measures the information about one sensor that is shared by another sensor in %$E_1$, where $E_1$ is 
the set of signals in $D$. It is seen that $\omega$ changes as soon as a sensor fault occurs, because the faulty signal is not present in the reference or other sensor signals.

\begin{table}[t]
    \centering
    \begin{tabular}{l}
        \toprule
\textbf{Algorithm 1: Non-faulty Data Collection for Damage Detection}\\ %%%%% [5060, algorithm 1] [5093, Table 1]
        \midrule
\textbf{Input:} a neighborhood with a bounded degree, \\ 
\hspace{8mm} $t \longleftarrow $ transmission at a time slot\\
\textbf{Step1:} \textbf{for} $i=1$ to $m$ \textbf{do:}\\ 
\hspace{8mm} node $i$ acquires data and buffers it \\
\textbf{Step2:} $i$ transmits its data to its neighboring node $j$\\ 
\hspace{8mm} \textbf{for} $t=1$ to $n$ \textbf{do:}\\ 
\hspace{11mm} \textbf{for} each node $i$ \textbf{do:}\\ 
\hspace{13mm} \textbf{If } $i$ has data not forwarded to $j$ \textbf{then}\\
\hspace{15mm} transmits a new data to its $j$ in $t$\\
\textbf{Step3:} call Algorithm 2  \\
\textbf{Step4:} data aggregation (extracting frequency sets, compute $\Phi^k$)\\
\hspace{8mm} compute final $\Phi$ \\
\textbf{Step5:} transmit the final $\Phi$ toward the BS\\
\textbf{return} health status of every sensor location\\
			\specialrule{1pt}{1pt}{1pt}
    \end{tabular}  
\end{table}

 %Upon receiving an event candidate, each sensor node evaluates its own currently available basic events and event candidates and depending on the
%settings of the distributed aggregation algorithm sends an acknowledgement
%to the originating node. Similar to the local event detection layer, the pa-rameters of whether to acknowledge an event candidate or not depend on the
%application and need to be carefully established before an actual deployment.
%If the sensor node that originally broadcasted the event candidate within
%its neighborhood receives enough positive replies within a certain period of
%time, it may safely regard the event candidate as a confirmed event, and thus
%send it to the base station

We use a joint Gaussian distribution based correlation model. Multivariate Gaussian distribution has been used to accurately model the correlation of many types of signals in literature \cite{3015}%[23]
. Each signal is broadcast to sensors in $D$, where $i$th sensor signal  $y_i^t \in y_{{D}}^t$ and $j$th sensor signal $y_j^t \in y_{{D}}^t$,  $i,j \in D$
 and %$N \subseteq D$ and 
$D \subset P$. For simplicity, $y_i^t $ as $u$ and $y_j^t $ as $v$ are denoted hereafter.

Hence, it would be worth considering how to find joint probability density between two signals. The statistical dependency/independency between the two Gaussian distributed time signals $u$  and  $v$  can be expressed in the form of the joint probability density $p(u, v)$ of signals:% as follows: % 
\begin{equation}% 
\begin{array}{l}
 p(u,v) = \frac{1}{{2\pi {\tau _u}{\tau _v}\sqrt {1 - {\rho _{uv}}} }}e - \frac{1}{{2(1 - \rho _{_{uv}}^2)}}\quad \quad \quad  \\ 
 \quad \left[ {{{\left( {\frac{{_{u - {\mu _u}}}}{{\tau _u^2}}} \right)}^2} - 2{\rho _{uv}}\frac{{(u - {\mu _u})(v - {\mu _{_v}})}}{{{\tau _u}{\tau _v}}} + {{\left( {\frac{{_{v - {\mu _{_v}}}}}{{\tau _v^2}}} \right)}^2}} \right]{\kern 1pt} 
 \end{array}
\end{equation}
where  $\mu _u$, $\mu _v$, $\tau _u$, and $\tau _v$ are the means and the standard deviations of the signals $u$  and $v$, respectively.  $\rho _{uv}$ is the correlation coefficient between the two signals. It is given by: 
\begin{equation}% 
{\rho _{uv}} = \frac{{E\left\{ {(u - {\mu _x})(v - {\mu _y})} \right\}}}{{{\tau _u}{\tau _v}}}{\rm{   }}  
\end{equation} 
The correlation coefficient can also sometimes be used to determine if two signals are statistically independent. On one hand, if $|{\rho _{uv}}| = 1$, there is a strong correlation between the two signals. On the other hand, if $|{\rho _{uv}}| = 0$, the two signals are not correlated. The correlation can be interpreted as a weak form of statistical dependency. In \cite{3011}, it is shown that two random variables, which are not correlated, can even so be statistically dependent. This is why we take the statistical dependency or independency. The product of the marginal densities $\rho _u$  and $\rho _v$ of the signals $u$  and $v$, respectively, is given by%stated as follows
:  
\begin{equation}
p(u,v) = p(u)p(v){\rm{ }} 
\end{equation}
If the expression in (11) is equal to the product of the marginal densities in (13), the signals are completely independent. %Fig. 2 illustrates the statistical dependency. 
	One possibility to quantify the statistical dependency between two signals is to calculate the MII of them, as follows:  
\begin{equation}
\omega (u,v,C) = \int \int {p(u,v)\log \frac{{p(u,v)}}{{p(u)p(v)}}} du\,dv{\rm{ }}  
\end{equation}
The base of the logarithm determines the units in which information is measured. (14) shows that if $u$  and $v$  are independent, $\omega$  becomes zero. 
%The base of the logarithm determines the units in which information is measured.
%${z_k}$ >>${h_o}$ 
A forward approach is to divide the range of $u$  and  $v$   into finite bins and to count the number of sampled 
pairs of ${h_o} = ({u_o},{v_o}), o = 1,2,\cdots,n $, falling into these finite bins. This count allows for approximately determining the probabilities, replacing (15) by the finite sum: 
\begin{equation}
{\omega _{bin}}(u,v,C) = \sum\limits_{a,b} {{p_{uv}}(a,b)\log \frac{{{p_{u,v}}(a,b)}}{{{p_u}(a){p_v}(b)}}} {\rm{ }} 
\end{equation}
where ${p_u}(a) \approx {n_u}(a)/n$ and ${p_u}(b) \approx {n_u}(b)/n$ are the probabilities based on the number of points ${n_u}(a)$ and ${n_v}(b)$ falling into the $a$th bin of $u$  and the $b$th bin of  $v$, respectively. The joint probability is $p_{uv}(a,b) \approx n(a,b)/n$   based on the number $n(a,b)$ of points falling into box nos. $a, b$. MII is non-negative and symmetric: 
\begin{equation}% 
\omega (u,v,C) = \omega (v,u,C) \ge 0{\rm{ }}  
\end{equation}
The MII for all possible combinations of sensor outputs $y_r $ and $y_s $ (except $r=s$, %where 
$i=1,2,\cdots,r, j=1,2,\cdots,s )$ is computed, which leads to an $\omega$-matrix for all combinations of $r$ and $s$. The basic 
idea is that the MII changes when a sensor fault ${f_r}$ is present. Suppose that it is in the $r$th channel or index: 
\begin{equation}
{\tilde y_r} = {y_r} + {f_r} 
\end{equation}
This fault appears only in the $r$th channel. Thus, we should expect that all combinations with index $r$ should show a reduction of $\omega$. This allows us to localize the faulty sensor. One or more faulty sensors can be simultaneously detected in the same way. One possibility to visualize the faulty sensor is to use the relative change as a sensor fault indicator $\lambda _{{y_r}}^\omega$: 
\begin{equation}
\lambda _{{y_r}}^\omega  = \frac{{|{\omega _{{y_r}}} - {\omega _{ref}}|}}{{{\omega _{{y_r}}}}}{\rm{  }} 
\end{equation}
where $y_r$ is an actual data set and the lower index \emph{ref} is one reference data set. The method based on MII is able to detect sensor faults in different combinations of them.

\vspace {-3mm}
\subsection{Algorithm 2: Faulty Sensor Detection}
\vspace {-1mm}
Under centralized detection, the BS handles the damage and faulty sensor detection process. In each decision cycle, the BS makes a decision about the faulty sensors, solely based on the $k$ most recent signals received from each sensor. The BS computes the MII for each signal, and chooses the signal with the maximal independence for fault detection. This detection is not suitable for resource-constrained WSNs. For example, if each sensor needs to send all its signals to the BS (where each sequence of signals  or raw natural frequencies can be from $X$0kb to $X$000kb, $X=1,2,\ldots$), the centralized WSN may not be able to operate for a given period of time. %Particularly, 
In a large-scale WSN deployment, the situation becomes serious. %suffers from many problems. 
After capturing data at high frequency in SHM, sensors should reduce data before transmission. 

In contrast, the faulty sensor detection (see Algorithm 2) can execute in a distributed manner where each sensor makes a decision on the collected signals locally, as described earlier% in our WSN framework
. %A sensor node decides whether it is faulty based on the signals from the neighbors. 
In the algorithm, if the local decision on a sensor's signals, $\lambda _{{y_r}}^\omega >0.5$, the sensor is faulty. This means that MII is high on the sensor's measured signals.  
The distributed method only requires neighbors to be synchronized. In addition, the detection is almost immediate and online, since a sensor does not need to wait for the signals from sensor nodes at more than one hop away. Moreover, the detected faulty signal set is not forwarded toward the BS; thus, the communication cost is relatively low. The energy cost becomes lower.

\begin{table}[t]
    \centering
    \begin{tabular}{l}
        \toprule
\textbf{Algorithm 2: Distributed Faulty Sensor Detection}\\ 
        \midrule
 Decision: ($\lambda _{{n_r}}^\omega  \le 0.5$: non-faulty), ($\lambda _{{n_r}}^\omega  > 0.5$: faulty) \\ 
\textbf{for} each sensor $i \in N$ where $N\subseteq D$ \textbf{do:}\\ 
\hspace{2mm} //set initial decision as a uniform distribution \\
\hspace{4mm} $(\lambda _{{n_r}}^\omega)_i  \longleftarrow 0$ // each sensor $i$ is non-faulty\\
\textbf{loop}\\
\hspace{2mm}\textbf{for} each sensor $i$ \textbf{do:}\\ 
\hspace{4mm} $(\lambda _{{n_r}}^\omega)_{j(neighbor)}  \longleftarrow $  receive neighbor $j$th decision\\
\hspace{4mm} samples $\longleftarrow N$ samples from $(\lambda _{{n_r}}^\omega)_{j(neighbor)}$\\

\hspace{6mm}\textbf{for} each sample $h$ \textbf{do:} // $h$ is an index\\ 
\hspace{8mm} $R_h  \longleftarrow $ non-faulty sensors \\
\hspace{8mm} $F_h  \longleftarrow $ faulty sensors \\
\hspace{8mm} $f_h \longleftarrow (R_h, F_h)$ //Eq. 3 \\
\hspace{2mm}\textbf{for} each sensor $i$ \textbf{do:} \\
\hspace{4mm}\textbf{if} $\lambda _{{n_r}}^\omega  > 0.5$ \textbf{then} \\
\hspace{6mm} $i$ marks/reports itself as a faulty sensor node \\
\hspace{4mm}\textbf{if} $i$ does not transmit the decision \textbf{then} \\
\hspace{6mm} $j$ marks/reports about $i$ as a faulty sensor node \\
 \hspace{2mm} call Algorithm 3 //  $i$th sensor signal reconstruction \\
\textbf{end}\\
			\specialrule{1pt}{1pt}{1pt}
    \end{tabular}  
\end{table}

MII does not rely on particular fault types. The algorithm 2 based on MII is able to detect different kinds of faults (as modeled before). However, it may fail to detect a node missing or failing. We provide  Appendix E for handling the node failure or node missing. 

%However,  Wireless sensor debonding (i.e., loosens the attachment to the structure so that a sensor cannot properly capture the vibration signal); sensor deterioration (precision degradation) depending on the wireless sensor types. For example, one sensor can be more vulnerable than another, e.g. strain gauges usually have a shorter life time expectation than accelerometers. Some common sensor faults can be detected, including short fault--a single-sample spike in sensor signals, random fault-- a longer duration of noisy or random signals, constant fault-- anomalous constant offset signals, etc. %drift fault  shifts of mean and/or variance of the sensor signals and losses of sensitivity lead to the reduction of sensor signalsˇ variance, and 
%and loss of sensitivity -- sensor signalsˇ variances are reduced often due to the aging of the sensor. %Malicious attacks may also be classified as a special type of fault that may be detected. An attacker introduces biases to sensor  signals through packet injection/modification% or compromised node
%%Unlike other type of faults, the biases in malicious attacks is hard to mathematically modeled due to different attacker behaviors.

\vspace {-2mm}
%%%%%%%%section%%%%%%%%%
%%%%%%%%section%%%%%%%%%
\section{Faulty %Sensor's 
Signal Reconstruction}
\vspace {-1mm}
In this section, we propose a Kalman Filter technique (KF) for faulty wireless sensor's signal reconstruction. 
The KF has received extensive attention to describe the recursive solutions of predicting state variables for linear systems \cite{3026}. We consider it, as it can generate the best estimation if the optimal filter is linear among all the linear observers, because it minimizes the error covariance. %Since the Kalman filter only works for linear system, the extended Kalman filter was developed to deal with nonlinear system in which the system has been linearized. 
In ACMS engineering domains, the KF has been  studied for %SHM and 
on-line damage detection \cite{2010}. Here, we utilize it for faulty sensor signal detection. 

%\begin{figure}[tb]
%\begin{center}
%\includegraphics[scale=0.6]{Figure-StateSpaceKL-Process}
%\end{center}
%%\vspace {-5mm}
%\caption {Graphical representation of the state-space equation based Kalman filter.} 
%%\vspace {-4mm}
%%\vspace {2cm}
%\end{figure}

\vspace {-2mm}
\subsection{Kalman Filter in State Space Representation}\vspace {-1mm}
The description of KF is made with the help of the state-space representation of the structural system, as described in Section 4.2.
One radical concept of the KF is that the state estimation is recursively corrected by the actual physical system outputs. Then, using (6), the equation of motion for time discrete and time invariant cases are given as follows:\vspace {-2mm}
\begin{equation}
\begin{array}{l}
 {l_t} = {M_{t - 1}}{z_{t - 1}} + {K_{t - 1}}{u_t} + {\sigma _{t - 1}}\\ 
 {m_t} = {M_t}{z_t} + {K_t}{u_t} + {\sigma _t} \\ 
 \end{array} 
\end{equation}
$u_t$ is the excitation at a specific frequency at time $t$. $M_t$ and $
K_t$ are transition matrices. The signals $\sigma _{t-1}$ and $\sigma _{t}$  represent the measurement noises, respectively (refer to Fig. D in Appendix D for the state-space equation-based KF). When measuring the responses of a dynamical structural system by %wireless
 sensors, the actual signals produced by the sensors are contaminated by noise due to internal manufacturing defects, physical interference,  or external environmental effects. According to features of the KF, we assume that every measurement from the wireless sensors contains noise; thus, if the noise measurement is zero,
the KF collapses. Setting the mean of noise as zero is a common practice: $E[\sigma _t]=0$.  Noises are assumed to be independent of each other,
 and are normally distributed with covariance matrices, $c_v= [\sigma\sigma^T]$. % and $c_\sigma= \sigma\sigma^T$ 

\begin{figure}[tb]
\begin{center}
\includegraphics[scale=0.65]{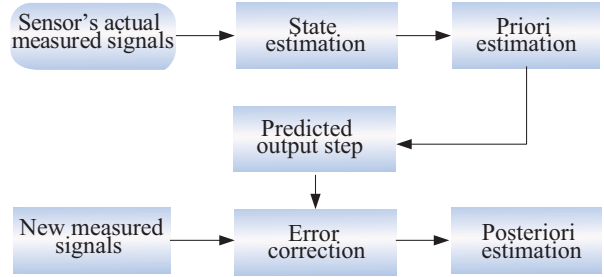}
\end{center}
\vspace {-4mm}
\caption {Sensor signal process flow under the KF.} 
\vspace {-4mm}
\end{figure}

The underlying  KF information is that KF is a recursive algorithm consisting of a loop (see Fig. D )
 which is passed through for each time instant $t$.   The estimation of the system state for $t$ is determined from the weighted average of the \emph{actual measured value} at time instant $t$ and the \emph{prediction of the system states} for this time instant. The weight factors of this average are determined from estimated uncertainties in each loop, which are also connected to the predicted system state and to the new measured value. The lower the uncertainty, the higher is the weight factor; i.e., Kalman gain ($K_t$). The uncertainty is calculated with the help of covariance matrices \cite{3027}. 

We shortly describe the faulty signal reconstruction process% based on [15]
. At first a \emph{priori} state estimate $P_k$ for the state vector $l_t^{pr}$ of the system are estimated \cite{3027}:\vspace {-2mm}
\begin{equation}
\begin{array}{l}
 l_t^{pr} = {M_{t - 1}}{z_{t - 1}} + {K_{t - 1}}{u_t} \\ 
 {P_k} = {M_{t - 1}}{z_{t - 1}}M_{_{t - 1}}^T + {c_v} \\ 
 \end{array} \vspace {-1mm}
\end{equation}
After getting the measured value $l_t^{pr}$, a \emph{posteriori} state can be estimated in the  correction step, see (21). For the \emph{posteriori}
 estimation, the difference between the measured and estimated signals are weighted by the Kalman gain factor $K_k$. %, see (26).
\vspace {-1mm}
\begin{equation}
l_t^{post} = l_t^{prio} + {K_t}[{m_t} - {M_t}l_t^{prio} - {K_t}{u_t}] = l_t^{prio} + {K_k}[{m_t} - m_t^{prio}]
\end{equation}
\vspace {-3mm}
\begin{equation}\vspace {-2mm}
{K_k} = {P_k}M_t^T{[M_t^{}{P_k}M_t^T + {c_v}]^{ - 1}}
\end{equation}
The \emph{priori} estimated error covariance $P_k$ in the prediction step is used to update the Kalman gain factor in (21) whereas $P_k$ itself is updated by the a posteriori estimated error covariance. The KF is used here based on the Matlab implementation, which delivers the optimum Kalman gain ($K_k$) together with the steady state error covariance matrix ($P_k$).

For the system state estimation using the KF, the process and the measurement noise covariance are determined. Since the system input is assumed to be unknown and the model uncertainty is high, the process noise covariates are set to high values. 
The values for the measurement noise covariance matrix will be determined a \emph{priori} by the first estimation of the measurement error ${m_t} - l_t^{post}$, as given in (21).  The overall process flow of the KF is illustrated in Fig. 5.

%%%%%%%%%%%%%%%%%%%%%%%%
\vspace {-3mm}
\subsection{Sensor Signal Reconstruction Algorithm}
\vspace {-1mm}

\begin{table}[t]
    \centering
    \begin{tabular}{l}
        \toprule
\textbf{Algorithm 3: Sensor Signal Reconstruction}\\ 
        \midrule
Step 1: Detect the faulty sensor by changes \\
\hspace{8mm} of indicator  $\lambda _{{n_m}}^{MI}$ (Algorithm 2) \\
Step 2: Recreate the state space model of the structure \\ 
\hspace{8mm} w.r.t. the sensor location (including faulty or \\ 
\hspace{8mm} nonexistent sensor) on the structure \\ 
\hspace{8mm} [see $M$ and $K$ matrices in Eq. (6)]\\
Step 3: Determine the process and measurements noise \\
\hspace{8mm} covariance matrices //(see Eq. (20)) \\
Step 4: Set the covariance value(s) of the assumed \\
\hspace{8mm} faulty sensor(s) to be high \\
Step 4: Solve the state space equation of motion through \\
\hspace{8mm} KF with the estimated state vector $l^{Post}_t$ (in signal \\
\hspace{8mm} correction step). The KF is driven by the input \\
\hspace{8mm} $u_t=0$ and the measured signals $y^i_t$ \\
			\specialrule{1pt}{1pt}{1pt}
    \end{tabular}  
\end{table}

If there is a sensor detected as faulty using Algorithm 2, the sensor signal reconstruction algorithm (see Algorithm 3) is used by the neighboring sensor nodes. 
The basic idea of the signal reconstruction is as follows. When a sensor signal does not correspond to the modeled system (monitoring its location) and it was erroneously assumed that this signal has a low measurements noise covariance, then the signals from the other sensors cannot be correctly reconstructed, and the difference between the measured and estimated signals will be high. If the value of the covariance for the faulty sensor signals is set as high, then the KF will reconstruct all the signals, including the incorrect signals, with the help of the other signals and the model. In this manner, it is possible to reconstruct more than one signal simultaneously. The number of signals that can be reconstructed rely on the number of neighboring nodes in case of  the distributed system, all of the nodes in the network in case of the centralized system, and on the quality of the model. For a better understanding of Algorithm 3, the procedure is broken into several steps.

By means of Algorithm 3, when just one sensor node does not work properly, it is possible to identify and reconstruct it only with the help of KF. For this purpose, Steps 2 to 5 are applied. Here, Steps 3 to 5 have to be calculated several times. The number of loops over these steps corresponds to the number of neighboring nodes in the case of the distributed WSN or all of the nodes in the case of the centralized WSN. 
 In each loop, the measurement variance of one sensor is set to be high. 
In addition, we attempt to detect a missing sensor and construct its signals that can be found in Appendix E.

%%%%%%%%%%%%%%%%%%%%%%%%%%
%%%%%%%%%%%%%%%%%%%%%%%%%%
\vspace {-3mm}
\section{Performance Evaluation}\vspace {-1mm}
\subsection{Simulation Studies}\vspace {-1mm}
\subsubsection{Methodology}
We conduct comprehensive simulations using MATLAB to evaluate \texttt{DependSHM} that includes the faulty sensor detection methods and signal reconstruction algorithm. We use real data sets %$^2$ [24]  
collected by the SHM system employed on the high-rise GNTVT \cite{1030} and a SHM toolsuite \cite{1518}.
%\footnotetext[2] {http://www.cse.polyu.edu.hk/benchmark/} % [8]. % to feed the data to the sensors. 
We use the data sets for the 100-sensor case in our simulations. % where the sensors are deployed through an optimal deployment method. 
We perform the WSN deployment via our WSN-based deployment scheme suggested in \cite{101}, which is supported by the ACSM engineering deployment methods \cite{501}.   %[6]% [23]
%in a deterministic manner [13] 
The simulation environment is a $450\times50$ sensing field regarding structural environment, e.g., bridge, building, aircraft. %An example of such sensor network is shown in Figure 3(a). 

The background data is simulated as vibration influenced by the 100 sensor locations in the field. 
A random Gaussian noise is added to all the data. The mean of the noises is zero, and the standard deviation is $10\%$ of the real signals. From the data sets, a set of data is used as \emph{reference data} to train the joint distribution, and another set of similar data is used for testing. %Note that 
A random Gaussian noise is added to all the data. The mean of the noises is zero, and the standard deviation is $10\%$ of the real signals. From the data sets, a set of data is used as \emph{reference data} to train the joint distribution, and another set of similar data is used for testing. %Note that 
The noise is present in both the data sets. Thus, the trained correlation model reflects the noises. 
In the distributed detection method, each sensor makes a decision based on signals received from neighboring nodes within $R_{min}$. 
%the communication radius. %Th iterative detection method starts with a random assignment of each sensor node being faulty or non-faulty. In each iteration, if a sensor node has changed its decision in the last iteration, it broadcasts the new decision to the neighbors. 
After a sensor receives a decision, it recomputes its MII and chooses to change its decision accordingly. %The search process terminates if no more change is made. % In each iteration, 50 samples are drawn at each sensor to evaluate the distribution. In the centralized random sampling method, 560 samples are randomly drawn from a uniform distribution and the one with the smallest mutual divergence value is used as the detection result.
The energy cost and routing models described in Section 3 are used for evaluation.

\begin{figure}[tb]
\begin{center}
\includegraphics[scale=0.25]{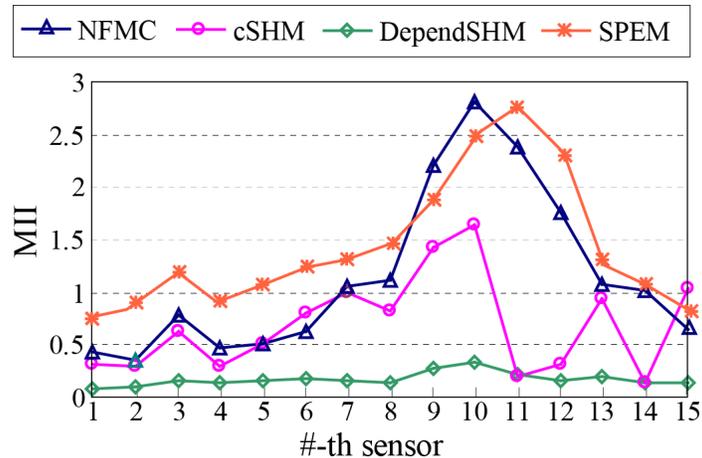}
\end{center}
\vspace {-3mm}
\caption {Performance of different fault detection methods: achieved MII under sensor faults.}
%; (b) the fault detection accuracy.}%; (c) observation on the actual mode shape curvature vs. mode shape curvature in cases of faulty sensor and recovery (signal reconstruction); (d) %energy cost in different measures analyzed by five rounds of monitoring; (e) fault detection ability vs. the number of sensors.} %\label{fig2}
\vspace {-4mm}
\end{figure}

%The maximum energy cost of a candidate location assignment depends on the routing protocol used by the data collection application. This falls into the domain of power aware routing (see section IVB). %Five general power-aware routing models were proposed in [19], and some other results can be found in [11][20]. 
%%The intuition behind is usually to route packets through the nodes that have higher residual power. 
%The \emph{maximum} energy cost by a node $i$ to send data correctly to a node $j$ is evaluated by the following model [16]:   
%\begin{equation}  
 %e_s (R) =  a R^{\alpha }+b%  
%\end{equation}
%which can also be normalized as: $ e_s (R) = R^{\alpha }+c$. $\alpha$ is an exponent parameter in [2, 4], $R$ is the communication range, and $c$ is a small constant comparing with $R^{\alpha}$. The energy cost for packet receiving is given by $ e_r (R) =c$ %We set $\alpha$ = 2 and $c$ = 4500
 %[16].
%%General power-aware routing models can be found in [7][21];  %The study of different routing protocols is out of the scope of this paper; for illustration purpose, we use the shortest path routing model [7].
%for evaluation purpose, we use the shortest path routing [7].

For comparison, we  implement other three schemes, including SPEM \cite{501} and NFMC \cite{523}. We compare their performance with  \texttt{DependSHM}.
We consider two schemes for observing the performance of  our fault detection and tolerance methods: i) distributed fault detection under localized data processing (\texttt{DependSHM}); ii) centralized fault detection under localized data processing (\texttt{cSHM}).
%;  iii) centralized fault detection under centralized processing data processing (\texttt{cSHM-LP}). 
SPEM is a WSN deployment method for SHM that nicely explains CS requirements and is verified on the GNTVT. It adjusts the quality of sensor locations to better fit WSN requirements. It is a centralized data processing method. We intend to verify its performance under fault detection and tolerance support and compare with \texttt{DependSHM}.
 NFMC is our preliminary fault detection and tolerance  WSN-based SHM scheme, which is based on the natural frequency extraction and matching.% in cluster. 

\begin{figure}[tb]
\begin{center}
\includegraphics[scale=0.75]{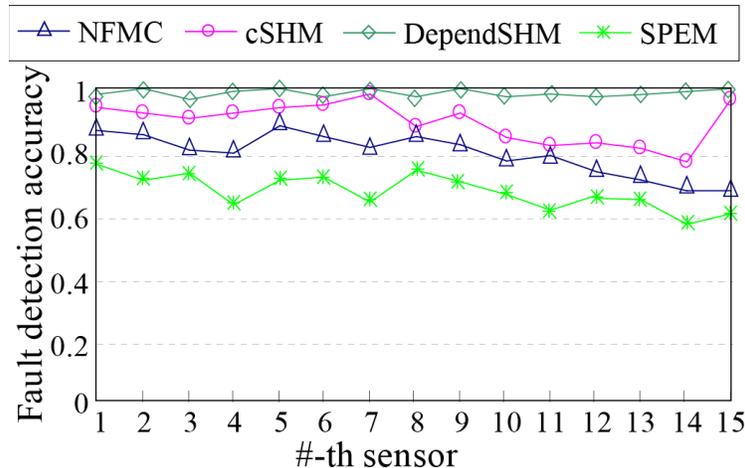}
\end{center}
\vspace {-3mm}
\caption {Performance of different fault detection methods: the fault detection accuracy.}%; (c) observation on the actual mode shape curvature vs. mode shape curvature in cases of faulty sensor and recovery (signal reconstruction); (d) %energy cost in different measures analyzed by five rounds of monitoring; (e) fault detection ability vs. the number of sensors.} %\label{fig2}
\vspace {-4mm}
\end{figure}

Using simulation results, we compare \texttt{DependSHM} with them in several aspects under the fault injections;: i) fault detection accuracy; ii) dependability in terms of detection ability and mode shape ($\Phi$) recovery, etc.,  and iii) energy cost of the WSN. %under distributed fault detection with localized data processing  in our scheme with cluster-based fault detection in [39].
Here, the \emph{detection ability} is the rate that is calculated by the percentage of successful faulty sensor detection to the percentage of the amount of the sensor fault injection. This includes both the false positive and false negative occurrences.
Here, false positive cases are recorded as an \emph{undamaged location} of the structure is identified as a \emph{damaged location}, and false negative cases are recorded as a \emph{damaged location} of the structure is identified as an \emph{undamaged location}.
 
%We also compute the false positive rate of \texttt{DependSHM}.

 %centralized fault detection under centralized processing (CFD-CP)
 
\subsubsection{Results}  
 In the first set of simulations, we implement all three schemes under the sensor fault injection (through modifying a number of sensors' signals randomly in the data sets). A fraction of the sensor nodes is randomly selected and the modified faulty signals are fed into their acquisition modules% as faulty sensors
. % by the values in the datasets. 
We vary the number of faulty sensors from $15\%$ to $25\%$. Each sensor node broadcasts its readings towards the neighboring sensor nodes. Each of the faulty readings is replaced by a random number independently drawn from a uniform distribution in the deployment field (0, 450).
 %Such a fault model is selected since it yields uncorrelated data in the same magnitude as the collected signals in practice. %the real environment. 
%
Fig. 6 shows MII achieved by  the four schemes. Out of them, \texttt{DependSHM} achieves the smallest value, followed by cSHM method. SPEM method performs poorly, since it requires centralized data processing and shows a significant amount of data packet loss during transmission. Due to heavy data losses, its performance on the MII is low. Nevertheless, NFMC still outperforms SPEM in many sensor fault detection cases. %The random sampling can usually find better results. In addition, the mutual divergence found by the distributed simultaneous is unpredictable at best. The method may fall into different local optimal given different random starting points. The algorithm of the centralized ran-dom sampling suffers from the same problem

Fig. 7 depicts the fault \emph{detection accuracy}, which is computed as accuracy = (true positive + true negative)$/$all. The detection accuracy in \texttt{DependSHM} is about $98\%$, which outperforms others. % two methods. 
In SPEM, the detection accuracy is poorer (less than $80\%$) than that of others, while it is from 75\% to 85\% in NFMC.
One major cause is that peak natural frequency signals used in NFMC and \texttt{DependSHM} achieve higher MII. However, we experience that the fault detection accuracy rate becomes lower in NFMC and SPEM than in \texttt{DependSHM} and cSHM, as the number of faulty sensor nodes in the WSN increases.
 
%Yet the distributed collective method is still able to find the true faulty sensor set even when 30% of the sensors are faulty. Note that when the number of faulty sensors is 7, the centralized random sampling method achieves higher detection accuracy than the distributed simultaneous method despite have higher mutual divergence value. This is because mutual divergence does not directly measure the number of faulty sensors. It instead measures how much the faults diverge from the real value. It is not uncommon that a faulty sensor exhibits larger divergence than the average divergence of several other faulty sensors.

%%%==================

\begin{figure}[tb]
\begin{center}
\includegraphics[scale=1.5]{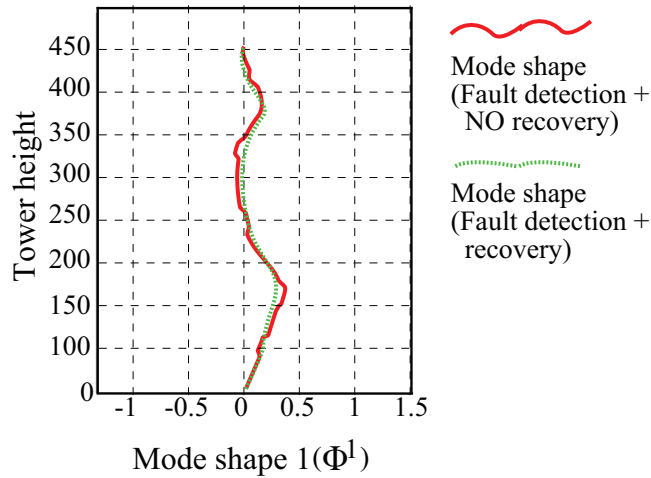}
\end{center}
\vspace {-4mm}
\caption {Observation on the actual mode shape curvature vs. mode shape curvature in cases of faulty sensors and recovery (signal reconstruction).}
\vspace {-4mm}
%\vspace {2cm}
\end{figure}

\textbf{Dependability Verification.} We next observe the structural health condition using WSNs in simulations. We estimate mode shape ($\Phi$) curvature under sensor faults and the signal reconstruction of the faulty sensors% in \texttt{DependSHM}
. We recover the first $\Phi$ (see Fig. 8) of the simulated structure with  100 locations, which cover up to 450 meters of the structure. $\Phi$  is extracted, 
%%%using professional vibration analysis software ARTeMIS Extractor, 
based on sensor collected signals in \texttt{DependSHM}. We can see the impact on the health status, in which the actual mode shape is distorted under the sensor faults, which is successfully recovered by the corresponding sensor signals' reconstruction. This implies that, if there is no appropriate faulty signal detection and tolerance methods, having successful monitoring operations will be difficult to achieve. Thus, a WSN-based SHM system without having such methods will not be dependable. 
More results and analysis of the \emph{performance of
 WSN-based SHM system dependability}  %(in terms of fault detection ability and structural event detection ability) 
in different schemes can be found in Appendix F.   

\begin{figure}[tb]
\begin{center}
\includegraphics[scale=0.28]{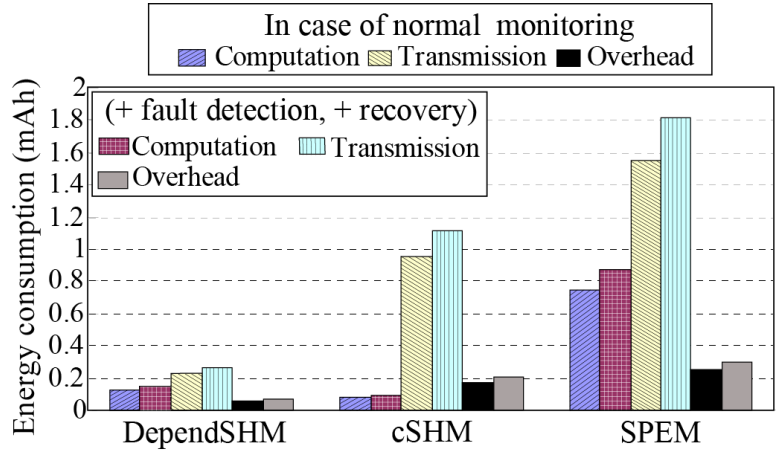}
\end{center}
\vspace {-4mm}
\caption {Energy cost in different measures analyzed by five rounds of monitoring.}
\vspace {-4mm}
%\vspace {2cm}
\end{figure}

\textbf{Energy Cost.}
We next observe the energy cost in the first five rounds of monitoring seen in Fig. 9 for \texttt{DependSHM}, cSHM, and SPEM schemes. We consider two cases: 
the amount of energy cost in the case of normal monitoring operation when there is no fault injection in the WSN and the WSN needs to provide health monitoring;
the amount of energy cost in cases of monitoring operations when there are the fault injection and recovery from the sensor faults through the signal reconstruction.  We calculate  the energy cost for computation, transmission, and overhead under both localized and centralized data processing.  We did not consider the energy cost for measurement, as we consider the same amount of energy cost for the measurement in all the schemes. We can see that the amount of energy cost for communication in SPEM and cSHM is very large compared to SPEM. 
The amount of energy cost in cSHM is seen to be around $60\%$ more than that of \texttt{DependSHM}, while it is  $90\%$ more than that of \texttt{DependSHM}. 

%\begin{figure}[t]
%\begin{center}
%\includegraphics[scale=0.8]{Figure9-ExpTestBed}
%\end{center}
%\vspace {-2mm}
%%\vspace {5.8cm}
%\caption{ (a) The SHM mote integrated by Imote2; (b)  twelve-story test structure and the placement of 10 SHM motes on it; (c) their deployment.} 
%\vspace {-2mm}
%\end{figure}

\begin{figure}[tb]
\begin{center}
\includegraphics[scale=0.4]{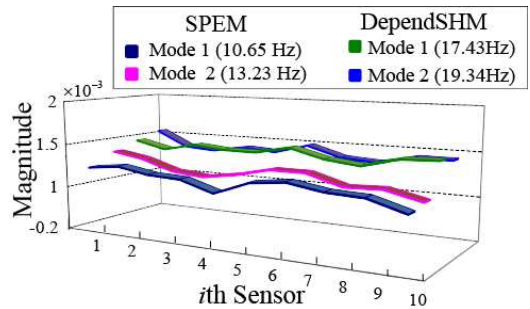}
\end{center}
\vspace {-4mm}
\caption {Global mode shape ($\Phi$) curvatures based on each sensor individual frequencies in SPEM and \texttt{DependSHM}.} %\label{fig2}
\vspace {-5mm}
%\vspace {2cm}
\end{figure}

\subsection{WSN Prototype System Implementation}
\subsubsection{Methodology and Wireless Sensor Platform} 
We validate our scheme by implementing a proof-of-concept system using the TinyOS on Imote2 platforms \cite{1515}. Our main objective is to verify i) the dependability and ii) the energy-efficiency of the system. 
We target the accuracy or successful $\Phi$  identification, because it can provide us with the answer, whether or not  a WSN-based SHM system is dependable in terms of various sensor faults. 
We employ 10 integrated Imote2s called SHM motes on a test structure (refer to Fig. G1 and Appendix G1 for more detail); an additional Imote2 is located 15 meters away as the BS mote, and a PC as a command center for the BS mote and data visualization. 
The test structure has 10 floors; at each floor, a mote is deployed to monitor the structure's horizontal accelerations. 
In the experiment,  $R_{min}$ is adjusted by the diameter of the structure, which is adjusted by estimating the height of the test structure and each floor.

\emph{Fault Injection.}  
To produce a sizable vibration response of the test structure, we collected the original data by vertically exciting the test structure using a magnetic shaker. 
%% removed
%Moreover, in order to observe monitoring results for a real structure, a dynamic change in the structure is made and is achieved by injecting enough excitation into the system to get measurements with a good signal-to-noise ratio. 
%
%
%In the sensor fault detection method, a sensor makes decision based on signals received from neighbors. In the algorithm, the initial decision distribution of each sensor is set to a uniform distribution. The background vibration signal is collected by Imote2 under measurement noise model.  A random Gaussian noise is added to all the signals. The mean of the noises is zero and the standard deviation is $10\%$ of the real signalsˇ standard deviation. 
%
%For the second set of experiments, the sensor fault is introduced by two cases
We inject the sensor faults in two cases: 
i) debonding fault %loosening the sensor attachment 
between the 5th sensor and the structure% and providing minimum power at the sensor 10 that the sensor can run about one hour
; ii) precision degradation fault during acceleration signal capturing by the 10th sensor. The sensors are expected to work properly but exhibit faulty acceleration measurements or decisions.
%(3) switching off the 10th sensor after 30 minutes of operation. Besides the sensor faulty signals, the sensor missing/failure is detected by the neighbors with the help of KL-KF changes as shown in Fig 9(a). 

\subsubsection{Experiment Results}
In the first set of experiments, we compute mode shapes ($\Phi$)  in the base-line structural system, when there are no damage events  and no sensor faults. These are computed by sensor initial identified natural frequency (presented in detail in Appendix G2). Fig. 10 demonstrates two mode shapes of the structure, captured by using the identified frequencies in both SPEM and \texttt{DependSHM} schemes. The results compare the exact  mode shapes  obtained by centralized WSN where the motes transmit their measured signals to the BS. On the other hand, in \texttt{DependSHM}, the final  mode shapes provided by the each mote are combined at the BS. The errors between two processes are analyzed. 
The accuracy of mode shapes  identification in SPEM is at least 13\% lower than the accuracy in \texttt{DependSHM}.
It is found that \texttt{DependSHM} has around $16\%$ better accuracy than that of SPEM under topology 2 (as shown in Fig. 3).
 The global  mode shape computed at the BS assembles all of the sensors' final results. Note that in the  mode shape assembling,  mode shapes from different motes correspond to the slight difference in the set of natural frequencies. 
%In this experiment,  mode shapes corresponding to the natural frequency set {12.85Hz, 13.55Hz, 14.67Hz, 15.86Hz} are obtained. 

Fig. 11a shows experimental fault detection results. Remarkable changes in signals of the 5th and 10th sensors and some of their neighboring nodes are detected. The MII changes in both of the 
sensor fault cases can be seen in Fig. 11b. Some of the neighboring  nodes, e.g., 4th, 6th, 9th, and so on have also provided an extent of change in their MII. This is because their signals have also been partially affected by the fault injection. The corrupted/faulty signals of a sensor (e.g., the 5th) are reconstructed (details  performance analysis on the signal reconstruction can be found in Appendix G3).

The energy cost analysis of the experimental WSN is provided in Appendices G4. We find that, in the case of faulty sensor detection and signal reconstruction, \texttt{DependSHM} consumes a small amount of energy in computation with a slight overhead, which is 5\% to 8\% of the total energy cost in each round, Meanwhile, it saves  a significant amount of energy for communication (which is at least three times when compared to its counterparts).

%The drift in the measured signal (red line) is corrected by the estimated signal (green lines). 
%We observe the mutual independence under the fault injection at the 5th sensor. 
%

%Furthermore, as shown in Fig. 13b, 
%SPEM achieves some amount of MII at almost  all of the sensor locations, even when the sensors are not faulty; the amount of MII is the largest in many sensor locations in the NFMC method. For example, sensors from 1st to 3rd and 6th and 9th show some MII. It may be considered as an ``abnormal signal". It may let neighboring sensor nodes fail in detecting fault and damage. 
%In the \texttt{DependSHM}, when sensor nodes process data locally, the small value in the MII is achieved, ranging from 2$\%$ to 4$\%$, and they are not considered faulty. The MII provides the best value, when there is a remarkable change in  the sensor measured signals, i.e., the 5th sensor and 10th sensor are faulty. This reveals that there is a better accuracy of fault detection in \texttt{DependSHM} in practice, compared to others.

\textbf{Dependability verification: Identification of what exactly happened in the structure. }
When computing the  mode shapes, we should notice that the signals of the faulty sensors contribute to the global  mode shape computation. Thus, the  mode shape values corresponding to the failed sensors are changed drastically. If there is a missing sensor, the mode shape result is affected. 
%(see Fig. 9a.
 Considering faulty signals, missing signals (in the case of sensor missing), or irregular signals (may be due to the damage), the  mode shapes will be affected. But we need to know exactly what happened in a WSN-based SHM system so that we can realize whether WSN-based SHM is dependable or not.
%by the  detection. These values are estimated using the cubic spline interpolation [8]. 
%

\begin{figure}[tb]
\begin{center}
\includegraphics[scale=1.1]{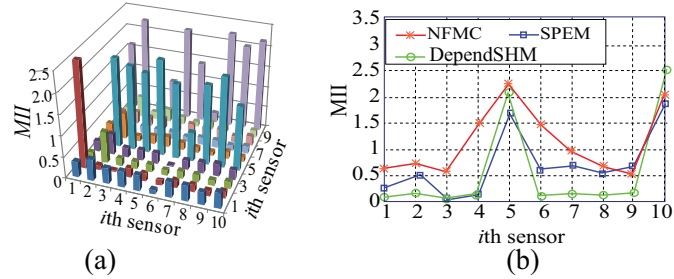}
\end{center}
\vspace {-2mm}
\caption {(a) Distributed sensor fault detection (5th sensor and 10th sensor are faulty);
 (b) MII achieved by the two detection methods under the sensor faults.} %\label{fig2}
\vspace {-2mm}
%\vspace {5cm}
\end{figure}

Now, we identify exactly what happened in the structure. Recall that there is a possibility of both sensor fault and damage occurrence at the same location.  
If there is a change in the signals with a single sensor only, the sensor may be faulty. If the change is present with multiple sensors, there is possibly damage.  However, if there is damage, it cannot be identified before the faulty sensor detection. 
As the fault injected, the 5th sensor should be faulty at a different time. As shown in Fig. 12a, during the SHM operation, at $T_d$ =5, the 5th sensor is detected as faulty, and at $T_d$=20, the 10th sensor is detected as faulty.  The changes in  the mode shape are computed at those time intervals.  

We inject structural physical damage through removing the plates on the 5th and 10th floors, since sensors located at these floors are faulty. They are not able to provide appropriate damaged information. 
We can see in Fig. 12b that the neighbors (4th, 6th, and 9th ) are able to detect an extent of changes (i.e., the presence of a damage) in the structure, 
where the slightly affected   mode shapes clearly appeared.

Under the same experiment and excitation setting, 
we further conduct experiments in which a faulty sensor signal is reconstructed by using our algorithm. %We can see the  mode shape results in Fig. ??.
The mode shape's  curvature is recovered significantly at the 10th sensor location, as shown in Fig. 12c. This means that there is possibly a damage, since the  mode shape is still slightly affected. However, the changes in  mode shape at the 5th sensor still remains and is slightly recovered at the neighbors. It provides the correctness of \texttt{DependSHM} and the dependability in WSN-based monitoring. If there was no recovery solution, the damage would not be identified and the changes at the sensor near the faulty sensor would not be discovered, which is the exact opposite of the 5th sensor cases. 

Further proof of the damage detection can be seen in Fig. 12d. Here, we replace the plate on the same floors. When the sensors wakeup and start monitoring, that time $T_d$=42. %Fig. 21 illustrates the  mode shape of baseline the structure, when there is no damage. 
In Fig. 12d, no remarkable change appears at the 10th sensor location. Distorted mode shape information at the neighbor locations is completely recovered. We can say that no MII appeared. In contrast, at the 5th location, mode shape remains unrecovered but there is no remarkable  mode shape curvatures at the neighbor sensors' location. It proves that the 5th sensor is surely faulty, while there was damage at the 10th sensor at this period of monitoring (but which is not clearly detected as the prior damage). 

Through an analysis, the quality of the faulty sensor signal reconstruction is about $92\%$ compared to the base-line results under the fault-free condition (as shown in Fig. 10, and TABLE G1 in Appendix G). 
The inference can be drawn from the above analysis that, in the presence of sensor faults, a damage can be successfully detected in \texttt{DependSHM}.

\begin{figure}[tb]
\begin{center}
\includegraphics[scale=0.4]{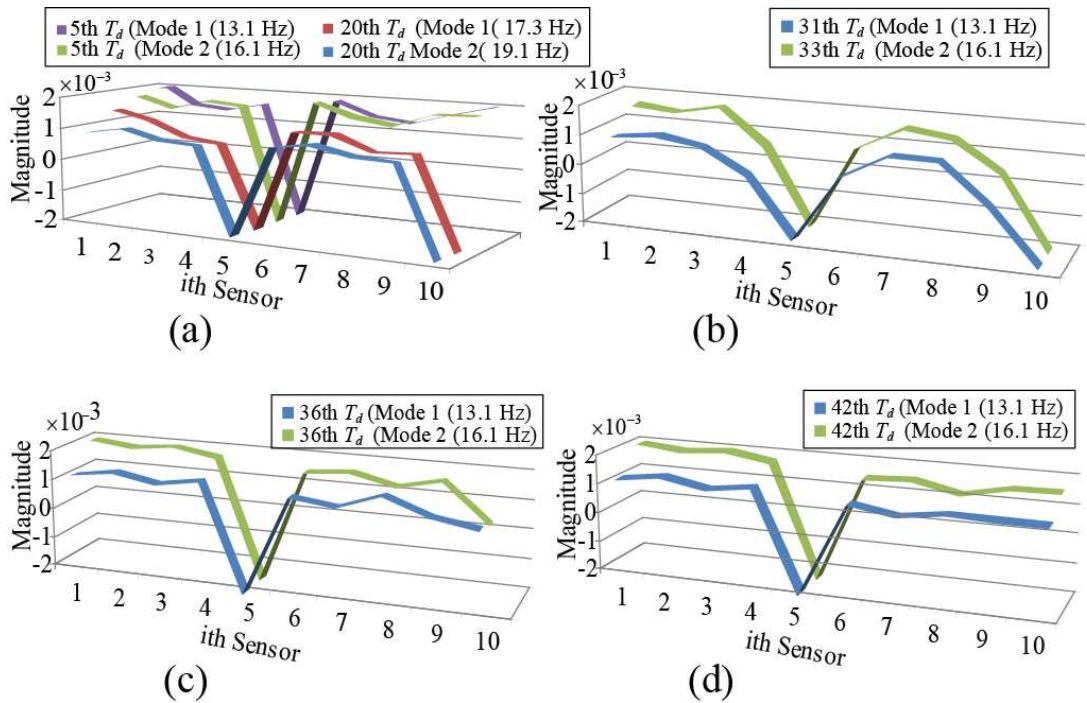}
\end{center}
\vspace {-2mm}
\caption {%Observation of 
The  mode shape's  curvature under sensor faults and recovery from the faults.} %\label{fig2}
\vspace {-3mm}
%\vspace {4cm}
\end{figure}

\section{Conclusion}
In this paper, we proposed a dependable WSN-based SHM scheme, \texttt{DependSHM}, by making the best use of resource-constrained WSNs for SHM and incorporating requirements of both engineering and computer science domains. \texttt{DependSHM} includes two complementary algorithms for sensor fault detection and faulty sensor's signal reconstruction. It is able to provide the quality of SHM in the presence of sensor faults automatically, which does not need any network maintenance for the fault detection and recovery, and does not consume significant WSN resources for the recovery. 
%We demonstrated the effectiveness of the \texttt{DependSHM} through extensive simulations and a prototype system implementation. %Both 
%
In the future, we plan to study decentralized computing architectures in WSNs, which can be integrated by the computing system issues and structural engineering system techniques in conjunction. Such an architecture is highly expected to reduce data traffic for data-intensive SHM and energy cost in WSNs.

\section*{Acknowledgment}
\addcontentsline{toc}{section}{Acknowledgment}
\small This work is supported in part by the National Natural Science Foundation of China under Grant Nos. 61272151, 61472451, 61402543, 61202468 and in part by ISTCP grant 2013DFB10070, in part by China postdoctoral research fund, in part by NSF under grants ECCS 1231461, ECCS 1128209, CNS 1138963, CNS 1065444, and CCF 1028167, and in part by HKRGC under GRF grant PolyU5106/11E and HK PolyU Niche Area Fund RGC under grant N\_PolyU519/12.

\bibliographystyle{IEEEtran}% 
\bibliography{refSRDStc}

\appendices

\section{Extended Energy Cost Model (cost($e_i$))} 
One of our important objectives is to minimize the energy cost of the network regarding various aspects, including the sensor fault detection and recovery, damage event detection. 
 Let $cost(e_i)$ denote the total energy cost of sensor $i$,  including measurement, computation, transmission, and overhead. The sensor $i$ has discrete power level that it can adjust it in ranges from $R_{\min}$ to  $R_{\max }$. In the beginning,  sensor $i$ adopts its minimum power level and then $i$ may dynamically increase it. 

We describe here how energy consumed in transmitting   %/receiving 
a packet. The maximum energy cost of $i$ depends on the routing protocol used by the data collection application. %This falls into the domain of power aware routing.
 
Consider a shortest path routing model \cite{101,501}; there is a path from sensor $i$ to neighboring sensor node or the BS $j$: $q=z_0,z_1 \ldots z_k$. 
Sensor $i$ propagates the data to them. We can find the $i$th hop sensor on each path and calculate the amount of traffic that passes along on the paths within each round of monitoring data collection ($T_d$, $d$=1,2,\ldots, $n$). Then, the $cost(e_i)$ can be decomposed into the following four parts:
\renewcommand{\theequation}{\thesection\arabic{equation}}
\begin{equation}% 
{\mathop{\rm cost}\nolimits} ({e_i}) = {e_{T}} + {e_{comp}} + e_{samp} +e_{oh}
\end{equation}
We describe these terms in the following:
\begin{itemize}
	\item  $e_T$ is the total energy cost %per bit
	for data transmission in a round of data transmission over a link between a transmitter and a receiver, where sensor $i$ uses its power level from a minimum to a maximum, but not beyond the maximum power. We use a standard energy cost model for calculating the packet transmission cost \cite{1525}. 

	\item 	
	%SEE for detail in DLACN TECT
	$e_{comp}$ is the energy cost for  processing data  locally, e.g., computing equation (6) in Section 4.2. If a sensor is allowed to transmit the raw vibration data to the BS directly, $e_{comp}$ would be very low.  The cost is mainly due to the onboard processor, such as a micro-controller, DSP chip, or FPGA \cite{3037}. These devices consume energy proportional to the number of processing cycles, as well as the maximum processor frequency $f$, switching capacitance  $\mu$,  and hardware specific constants $k$ and $\beta$, respectively \cite{3037}. We focus on the number of cycles taken for tasks, e.g., equation (6) and the amount of samples taken. The number of cycles required to perform a task on the amount of samples (denoted by $w$) are estimated according to the computational complexity $O(w)$, which  describes  how  many  basic  operations,  i.e.,  averages, additions, multiplications, etc., must be performed in executing the task. Given these parameters, the computational energy to complete a task can be calculated according to:
\begin{equation}
{e_{comp}} = O(w) \cdot \mu (\frac{f}{k} + \beta )
\end{equation}

	\item    $e_{samp}$ is the energy required for a sampling cost of $M$ data points; when sensors capture vibration signals, assuming a maximum $50\%$ overlapping, $M =( n_a/2+1/2) \cdot c_r$,  %$e_{samp}$ are the energy for sampling one unit vibration data and
 where $n_a$ and $c_r$ are the number of averages mainly for the purpose of noise reduction, that practically ranges from 10 to 20 and cross-correlation factor, respectively \cite{523,513}. We assume that $n_a$ and $c_r$ are set by fixed values on a sensor. 

\item $e_{oh}$ is any  additional overhead for some causes, e.g., fault detection and signal reconstruction, copying  data to a local buffer, and %other possible sources (e.g., 
network latency.

\end{itemize}

\section{Method of Extracting Local Mode Shape}
In Section 4.2, we have described the state-space model for structural mode shape computation ($\Phi$) at individual sensor. Here, we show a method to local $\Phi$ extraction and explain benefits of utilizing the extracted $\Phi$ over $f$ towards sensor fault detection and tolerance. 

\textbf{Definition B1 }[Natural Frequency]. \emph{Every structure has a tendency to vibrate with much larger amplitude at some frequencies than others. Each such frequency is called a natural frequency denoted by $f$. $f$ is an internal vibration signal characteristic of structure, and is different for different structures (such as from building to bridge, from indoor to outdoor).
In other words, it is defined as the number of times that a structural system oscillates (moves back and forth) between its original state and its displaced state when assuming there is no outside interference.}

%%%%%  Figure  %%%%%%
%\renewcommand{\thefigure}{\thesection.\arabic{figure}}
\renewcommand{\thefigure}{B1}
\begin{figure*}[tb]
\begin{center}
\includegraphics[scale=0.22]{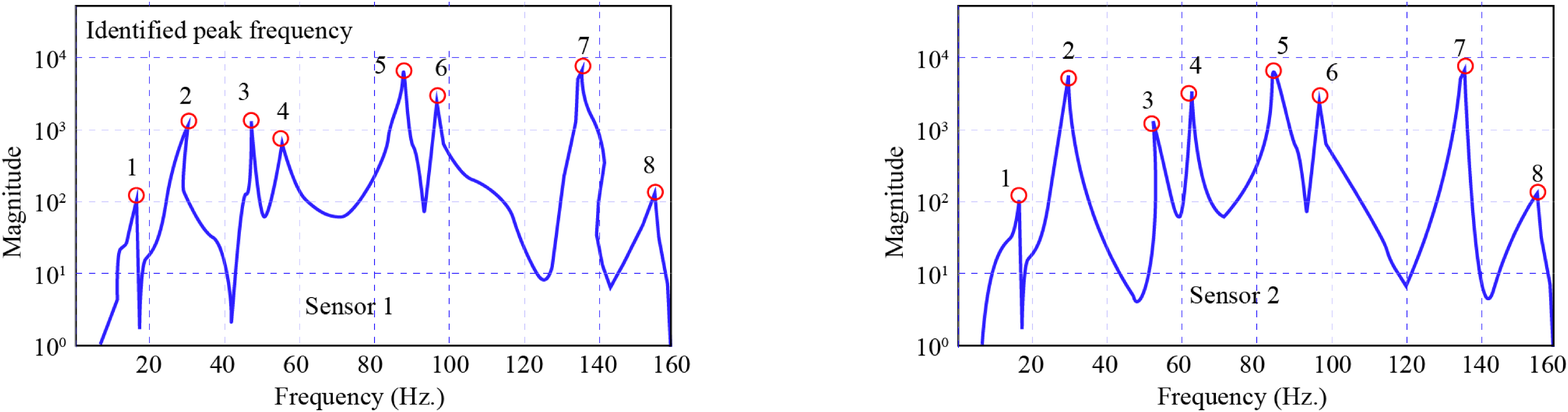}
\end{center}
\vspace {-1mm}
\caption{Based on acquired vibration signal characteristics, measured natural frequencies captured by sensor 1 and sensor 2 under manual input excitation on the structure, respectively. This shows the structural system oscillation (moving back and forth) between its original state and its displaced state, captured by the two in their vicinity.}
%\vspace {-3mm}
%\vspace {2.5cm}
\end{figure*}
%%%%%  Figure  %%%%%%

%
%%%%%  Figure  %%%%%%
\renewcommand{\thefigure}{B2}
\begin{figure*}[tb]
\begin{center}
\includegraphics[scale=0.22]{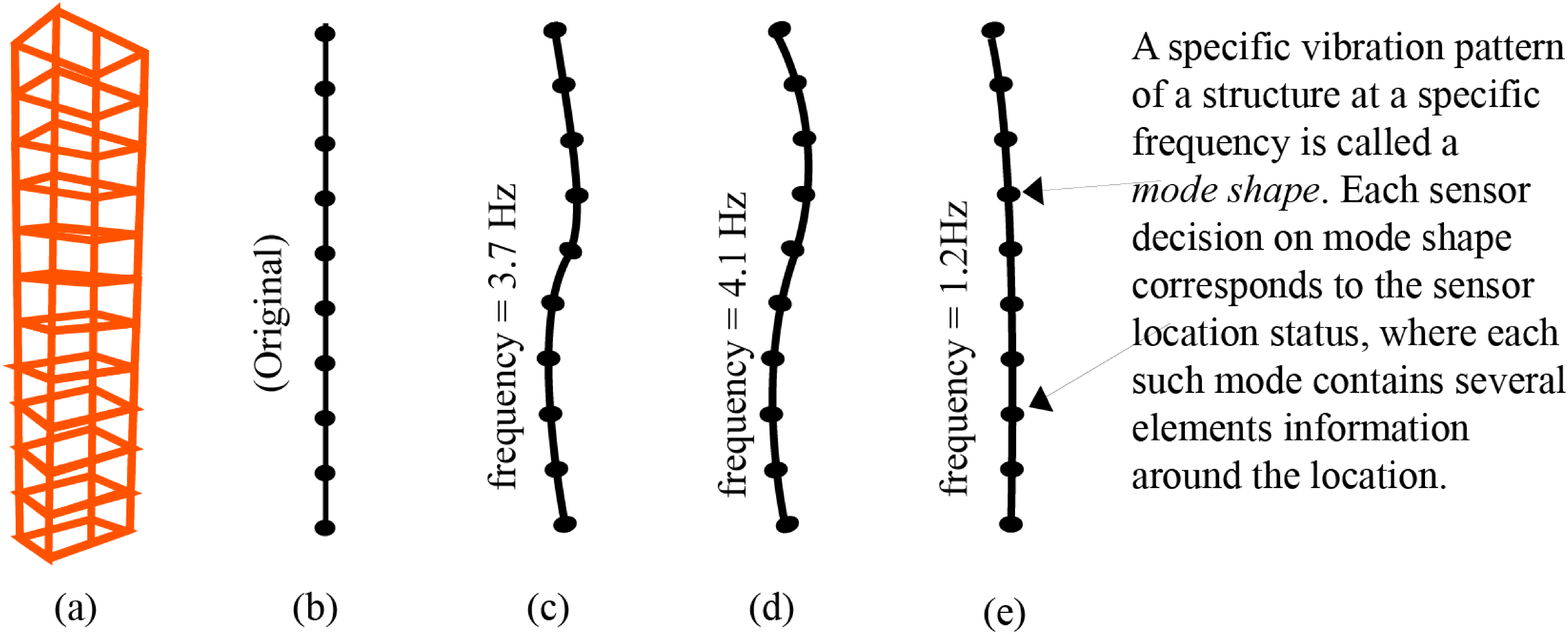}
\end{center}
%\vspace {-4mm}
\caption{ (a) The finite element model (FEM) of our designed physical infrastructure; (b) its original mode shape; (c)-(e) its three mode shapes: mode 1, mode 2, and mode 3. \textit{FEM} is a computer based
numerical model often used for calculating the behavior and strength of structural mechanics, such as vibration and displacement. }
 %\vspace {-6mm}
%\vspace {2.2cm}
\end{figure*}
%%%%%  Figure  %%%%%%

\textbf{Definition B2} [Mode shape]. 
\emph{When subjected to external forces, the response of a structure is conceptually similar to the response of a vibrating string or structural components such as a metal plate. Upon excitation, the vibrations are a combination of several harmonics (or at a specific frequency of vibrations), known as modes. Each mode deforms the structure into a particular spatio-temporal pattern known as a mode shape, denoted by $\Phi$.}

\subsection{Local Mode Shape by Each Sensor}

As the network modeled in Section 3.1, $m$ sensors are available for deployment on a structure, and they extract a total of $p$ mode shapes from the measurement of $m$ sensors. The corresponding natural frequencies and mode shapes are represented, respectively,  as follows: 

\renewcommand{\theequation}{\thesection1}
% 
%\begin{equation}% 
%{\textbf{f}}= [{f}^1, {f}^2, \ldots, {f^p}]
%\end{equation} 
%%
%\begin{equation}
%\Phi  = \left[ {{\Phi ^1}\;\;...\;{\Phi ^m}} \right] = \left[ {\begin{array}{*{20}{c}}
  %{{\phi _{11}}}& \ldots &{{\phi _{1p}}} \\ 
   %\vdots & \ddots & \vdots  \\ 
  %{{\phi _{m1}}}& \cdots &{{\phi _{mp}}} 
%\end{array}} \right]  
%\end{equation}

\begin{equation}
{\textbf{f}}= [{f}^1, {f}^2, \ldots, {f^p}]
\end{equation}
\renewcommand{\theequation}{\thesection2}
\begin{equation}
\begin{array}{l}
 \Phi  = \left[ {{\Phi ^1}\quad {\Phi ^2}\,\;.\;\;{\Phi ^m}} \right] = \left[ {\begin{array}{*{20}{c}}
   {\phi _1^1} & {\phi _2^1\;\;.} & {\phi _p^1}  \\
   \begin{array}{l}
 \phi _1^2 \\ 
 \;. \\ 
 \end{array} & \begin{array}{l}
 \phi _2^2\;\;. \\ 
 \;.\;\;\;\;. \\ 
 \end{array} & \begin{array}{l}
 \phi _p^2 \\ 
 \;. \\ 
 \end{array}  \\
   {\phi _{1\;}^m\;.} & {\phi _2^m\;\;.} & {\phi _p^m}  \\
\end{array}} \right] \\ 
  \\ 
 \end{array} 
\end{equation}

where ${f^k} (k= 1,\ldots,p$) is the $k^{th}$ natural frequency, ${\Phi^k}$ is the mode shape corresponding to ${f^k} \cdot \phi _i^k (i = 1,\ldots,m)$ is the value of  ${\Phi^k}$ at the $i^{th}$ sensor. For example, Fig. B1 and Fig. B2 illustrate the first two sensors' natural frequencies and corresponding mode shapes of a physical structure, receptively, which are extracted from measurements of 10 deployed  sensors in our prototype system. 
In the experiment, vertical accelerations at all the given sensors are obtained, and $10\%$ noise is added to all measurements.  Under the artificial input excitation, the measured accelerations (the peak frequency pointed by $1, 2,\ldots$) at sensors 1 and sensor 2, respectively, refer to Fig. B1 and Fig. B2 (which are obtained by using network topology in Fig. 3).
%The identified frequency sets from two healthy sensor nodes $\rm{f_1=}$ can turn out to be as

\subsection{$\Phi$ over $f$ in Sensor Fault Detection}

The difference between $f$ and $\Phi$ can be observed by comparing (C1) with (C2) and Fig. C1 with Fig. C2. According to the ACSM theory, $f$ is not suitable characteristic for damage event detection due to several reasons: 

\begin{enumerate}[(i)]
	\item $f$ is not a sensitive indicator to damage event, where only severe damage event causes noticeable change on the set of $f$;

\item  Due to the global property, $f$ does not contain any spatial information, and thus localizing damage event is difficult, while damage event detection using $f$ is computationally inefficient; 
\item High frequency modes are more susceptible to additional noise than low frequency modes; 
iv) $f$ is susceptible to additional noise
\cite{513}; To improve the usability of  the $f$  to  detect damage event of  small  magnitude, high-frequency modes, which are associated with local responses, may be used. However, 
we argue that adopting $f$ is not suitable for WSNs considering WSNs' resource limitation; %t makes the method ineffective for WSN
\item  Importantly, a large set of $f$ is required to be sent  to the BS %in most applications 
(e.g., SPEM \cite{501}, NFMC \cite{523}); damage event detection is greatly affected if a portion of it is lost during transmission. 
\item $\Phi$ is directly linked to topology of the structure and $\Phi$ highly features the dynamics of the structure.
\end{enumerate}

On the other hand, it can be seen from (C2) that $\Phi$ has elements corresponding to each sensor, thus containing spatial information. $\Phi$ and its derivatives have been proven to be very sensitive features to detect damage event. It takes into account out-of-frequency-bandwidth modes of the structure, and is also applicable to a complex linear structure. 
This is why we target on $\Phi$ computation and observe the impact of sensor faults on $\Phi$. 
However, theoretically, $\Phi$ is a global parameter of a structure which means that, using sensor deployed on different locations of a structure, the same set of $\Phi$ may not be obtained. To mitigate this problem, we allows each sensor estimate $\Phi$ taking measurements about its vicinity (i.e., local structural response).

%Therefore, it can be suitable for WSNs considering computation, transmission, data sharing, and fault handling. Traditionally, the identification of mode shapes is centralized and requires the aggregation of the raw data from all of the deployed sensors. We allow sensors to compute mode shape locally and transmit the final mode shape to a center (also known as a BS) considering limited wireless bandwidth and energy, and transmission reliability.
 In this paper, we utilize the mode shape curvature method proposed by civil engineering to identify significant change (i.e., damage event) in the mode shape \cite{1079}. The mode shape curvature has high sensitivity to damage event. %and is able to detect damage%, even without using the references that identified by when the structure is healthy

%computing mode shape locaaly reduces  transmsision tens of KB data compared transmitting all sets of frequencies

%%%%%%%%%%%%%%%%%%%%%%%
%%%%%%%%%%%%%%%%%%%%%%%
\section{The Reason of Neglecting Damping}
In Section 4.2, we have described the space-space model for the structural response measurements by sensors. In (6), we have considered the matrices %% for C for damping
of  mass and stiffness coefficients of the various elements of the structure, but we have neglected the damping. 

Damping is neglected in the model (6), considering individual sensor measurement estimation.  In any case, (a) faults in the sensors will only be identified if they cause changes in the response of a greater magnitude than the errors in the estimated mode shape, and (b) the modes with low damping, having approximately real modes, will be  strongly excited. Thus, undamped mode shapes can be  accurately estimated by (6) at each sensor.

%%%%%%%%%%Section%%%%%%%%%%%%%%%%%%
%%%%%%%%%%Section%%%%%%%%%%%%%%%%%%
\section{The State-space-equation based KF for Signal Analysis}

\renewcommand{\thefigure}{D}
\begin{figure*}[tb]
\begin{center}
\includegraphics[scale=0.75]{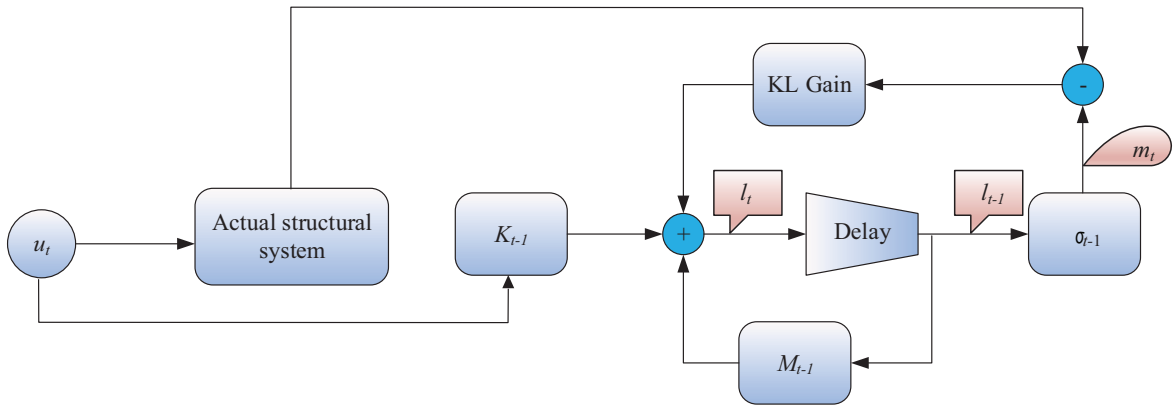}
\end{center}
%\vspace {-5mm}
\caption {Graphical representation of the state-space equation based Kalman filter that is used in sensor's faulty signal reconstruction.} 
%\vspace {-4mm}
%\vspace {2cm}
\end{figure*}

In Section 6, we proposed the Kalman filter (KF) technique for signal reconstruction. 
In this Appendix, a graphical representation of the state-space equation based KF is presented in Fig. D. This  equation is made with the help of the state-space representation of the structural system, as described in Section 6.1. 
In Fig. D, $u_t$ is the structural excitation at a specific frequency at time $t$. $M_t$ and $K_t$ are transition matrices. The signals $\sigma _{t-1}$ and $\sigma _{t}$ represent the measurement noises, respectively. When measuring the responses of a dynamical structural system by wireless sensors, the actual signals produced by the sensors are contaminated by noises due to internal manufacturing defects, physical interference,  or external environmental effects. According to features of the KF, we assume that every measurement from the wireless sensors contains noises; thus, if the noise measurement is zero,
the KL collapses. Setting the mean of noise as zero is a common practice: $E[\sigma _t]=0$.  Noises are assumed to be independent of each other, and are normally distributed with covariance matrices, $c_v= [\sigma\sigma^T]$. 
% and $c_\sigma= \sigma\sigma^T$.
The underlying  KF information is that KF is a recursive algorithm consisting of a loop,
 which is passed through for each time instant $t$.

%%%%%%%%%section%%%%%%%%%
%%%%%%%%%section%%%%%%%%%
\section{Missing Sensor Detection Method}

\renewcommand{\thefigure}{E}
\begin{figure*}[tb]
\begin{center}
\includegraphics[scale=1.2]{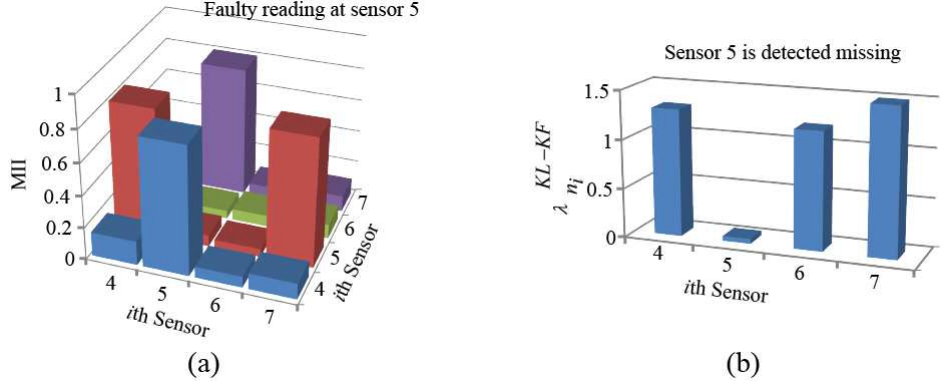}
\end{center}
\vspace {-2mm}
\caption {Missing or failed node detection in a WSN-based system: (a) an example of MII change in the sensor signals; (b) the detection result under the KL-KF.} %\label{fig2}
%\vspace {-4mm}
%\vspace {2cm}
\end{figure*}

In Section 5, we presented Algorithm 2 for faulty sensor detection. However, if a sensor is missing or is out of service during the monitoring operation, if the sensor cannot be reached because of communication constraint or failure, or there is an unknown reason, the algorithm cannot guarantee detection of such a sensor node. In order to detect these sensors, we apply a method of Kullback-Leibler divergence (KL) \cite{3025} between the measured and estimated sensor signals and update Kalman filter (KF) with KL, which can be used as a fault indicator $\lambda _{{n_i}}^{KL - KF}$ KL-KF (Kullback-Leibler-Kalman-Filter) for such sensors. Note that for a faulty sensor signal reconstruction, we will use the Kalman filter (KL) technique.

The symmetrized form of the KL between the probability distributions of one measured signal $(y_i)$ at time $t$ and with the KF estimated signal $\hat y_i$ is as follows: 
\renewcommand{\theequation}{\thesection1}
\begin{equation}%\vspace {-2mm}
KL = \frac{1}{2}\sum\limits_i {[{p_{y_i}} - {p_{\hat y_i}}]} {\log _2}\frac{{{p_{y_i}}}}{{{p_{\hat y_i}}}}
\end{equation}

where ${{p_{y_i}}}$ is the probabilities based on the number of points falling into the  $i$th bin. When the KL between two probability distributions is zero, the signals are identically distributed. The fault indicator is  defined as:
\renewcommand{\theequation}{\thesection2}
\begin{equation}
\lambda _{{p}}^{KL - KF} = \frac{1}{{{p_{\max }} - 1}}\sum\limits_{p = 1}^{{p_{\max }}} {K{L_p}},1 <p < p_{\max }
\end{equation}

If the faulty sensor $p$ is not used for the estimation of the $p_{max}$  sensor signals, then the KL distance between the measured and estimated signals will be minimal; otherwise, the distance will be higher. This is shown under the network topology in Fig. 3a. It can be seen in Fig. D that without sensor 5, the best estimation is possible, which clearly indicates the sensor is faulty. This method based on KF is able to detect a missing or failed sensor. Also, pure bias faults with the MII method are enhanced further by this KL-KF method. Thus, using KF-KL with the help of Algorithm 2, it can be guaranteed to detect the fault types that produce faulty readings.

We illustrate the justification of sensor fault identification method based on MII  (i.e., $\omega$) through Algorithm 2. In our real experiment, under the manual random excitation and 5th sensor removal, the 5th sensor is detected as faulty. As shown in Fig. E(a), the relative change in MII indicates the sensor 5 as faulty. Actually, the sensor was removed from the location, however, the sensor is detected as missing by $\lambda _{{p}}^{KL-KF}$, as shown in Fig. E(b). To guarantee a certain redundancy of information in each sensor data set, the initial frequency should be available for identifying the faulty vibration signal.  
Therefore, if one of the neighboring nodes is missing, the KL-KF divergence between the measured and estimated sensor signals can be used as a sensor fault indicator.

\section{More results of WSN-based SHM System Dependability}

\renewcommand{\thefigure}{F1}
\begin{figure}[tb]
\begin{center}
\includegraphics[scale=1]{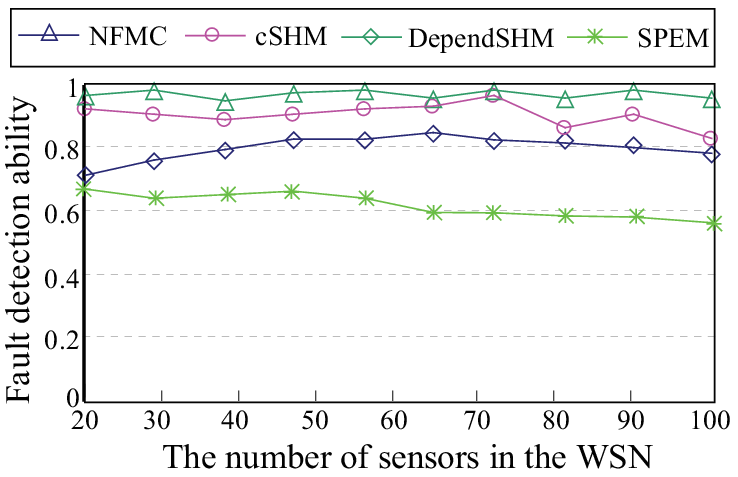}
\end{center}
%\vspace {-2mm}
\caption {Dependability verification: fault detection ability of different schemes.}
%\vspace {-2mm}
%\vspace {2cm}
\end{figure}

In Section 7.1.2, we have partly performed an analysis of the system dependability. In the analysis, we have used a combination of  true positive and true negative results in the sensor fault detection accuracy estimation. In this Appendix, we continue the analysis of the performance of the system dependability. We particularly consider the dependability of WSN-based SHM schemes as the ability of fault detection and the ability of structural health event (damage) detection of the schemes.

 At first, we discuss the detection ability of different WSN-based schemes. Fig. F1 demonstrates the fault detection ability of \texttt{DependSHM} and other schemes. We can see that the detection ability of \texttt{DependSHM} is much better than that of cSHM, NFMC, and SPEM. NFMC shows higher detection errors than \texttt{DependSHM}, even higher than cSHM. Looking into details of causes, we summarize the following observations under the random fault injection:

\begin{enumerate}[(i)]
	\item The same pick frequencies cannot be achieved in many neighborhoods or clusters in NFMC; 
	\item  One or more clusters are disconnected from the network, as one or more faulty sensors are isolated based on the natural frequency comparison (although it shows the good ability rate of fault detection in some clusters); 
	\item  The scheme is limited to the frequency matching based fault detection; 
	\item  NFMC fails to detect other types of faults; 
	\item The fault detection ability of SPEM is very low, due to non-faulty reading losses that results in a increased amount of faulty readings;
	\item  When attempting to recover from the faults, both SPEM and NFMC schemes require a significant amount of energy cost. 
\end{enumerate}

\renewcommand{\thefigure}{F2}
\begin{figure}[tb]
\begin{center}
\includegraphics[scale=1]{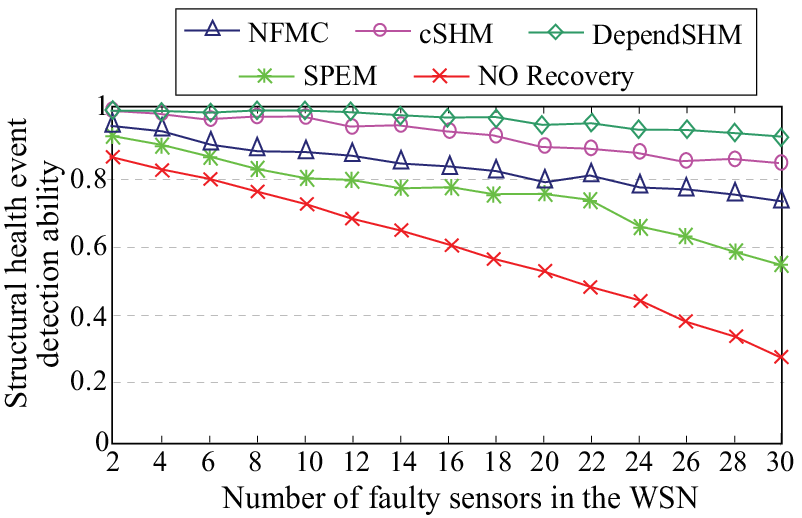}
\end{center}
%\vspace {-2mm}
\caption {Dependability verification: structural health event detection ability of different schemes.}
%\vspace {-2mm}
%\vspace {2cm}
\end{figure}

We next examine the system dependability in terms of the structural health event detection ability of a system. This can provide us an implication that how much a system can cope with sensor faults and what is the significance of addressing dependability issue in a system. We gather all the false positive and false negative cases appeared in the WSN-based SHM (achieved from a total of 50 simulation runs), and we get an average. Then, we calculate the structural health event detection ability rate as 1-(false positive rates + false negative rates). The results is depicted in Fig. F2. We also take into account the structural health monitoring under \emph{NO} recovery (a preliminary analysis has been done based on these results, as illustrated in Fig. 8). Here, we intend to find evidence that what exactly happens when there is \emph{NO} dependability option (fault detection and recovery) provided.

In Fig. F2, we can see the results, which shows that the structural event detection ability of \texttt{DependSHM} is between 93\% and 97.2\%, which greatly outperforms others. In SPEM, the detection ability under recovery from sensor faults tolerance algorithm is inferior (between 75\%  and 92\%) among all of the schemes, while it is between 74\% and 95\% in NFMC and 87\% to 95\% in cSHM. There can be various reasons that SPEM provides poor detection rate, including i) centralized decision making on the fault detection and tolerance (data losses on the fly is a factor), ii) application-specific sensor deployment, iii) natural frequency matching problem, and so on. In NFMC, the peak natural frequency signals used in the sensor fault detection and recovery, by which the actual mode shape curvature slightly distorted. This lead to a lower MII that results in a lower structural event detection ability.  As it can be seen in Fig. F2, the structural event detection ability 
becomes lower in NFMC and SPEM than in \texttt{DependSHM}
and cSHM, as the number of faulty sensor nodes in the
WSN increases.

From the results in Fig. F2, the structural event detection ability rate is around 65\% in a system with \emph{NO} recovery from sensor faults. It may make us surprised that the monitoring operations in a WSN-based SHM can be often meaningless  if there is no dependability option provided. From a deep observation, we have found evidence that faulty sensors can corrupt results of a  health event in a structural system without being detected. We have seen that measured signals introduced by some faulty sensors often identify its location as damaged (actually it is undamaged location). We also have found that some faulty sensor identify its location  as undamaged (actually the  location is damaged). There are a large number of such wrong diagnoses (false positive and false negative) that lead to a reduced structural event detection ability.

%%%Our technique addresses this problem
%%%by focusing attention to fault detections that correspond
%%%to very low probability events, which improves accuracy
%%%of fault detection from 5% to 65% on faulty runs, while
%%%maintaining a 5% false positive rate.

\section{More Details of the WSN Prototype System Implementation}
\subsection{Extended Detail of the Experimental Setup}
We validate  \texttt{DependSHM} by implementing a proof-of-concept system using the TinyOS  on Imote2 platforms \cite{1515}. Our main objective is to verify i) the dependability 
% (ability of fault detection and structural health event detection 
%accuracy of  $\Phi$  identification in the presence of sensor faults 
and ii) the energy-efficiency of the system. 
We target the accuracy or successful $\Phi$  identification, because it can provide us with the answer, whether or not  a WSN-based SHM system is dependable in terms of various sensor faults. 

%The smart sensor is able to computes the  $\Phi$  and evaluating self-fault status or neighborˇs fault status. 
The Imote2 (IPR2400) is an advanced wireless sensor platform (off-the-shelf), offering  sufficient processing capability and communication resources to locally and continuously monitor vibration characteristics under intensive conditions. Its main board combines a low power PXA271 XScale processor with an 802.15.4 radio (CC2420) and an antenna using 2.4 GHz. The major limitation with it is the energy.  % It also offers 256 KB of integrated SRAM and 32 MB of external SDRAM.  

We employ 10 integrated Imote2s called SHM motes on a test structure, as shown in Fig. G1; an additional Imote2 is located 15 meters away as the BS mote, and a PC as a command center for the BS mote and data visualization. 
The test structure has 10 floors; at each floor, a mote is deployed to monitor the structure's horizontal accelerations. 
Each mote runs a program (implemented in the nesC %programming
 language) to process the acceleration data acquired from on-board accelerometers. 
%Acceleration data is collected by a digital sensor board at a specific sampling frequency.
 %Digital acceleration data, acquired within frames of %$M=%
%2048 points, is then stored in the local memory for each period of monitoring. 
%On-board digital signal processing is performed on the stored data to obtain a set of intermediate parameters, which are then transmitted to the BSto complete the algorithm. 
%The accelerometer has a resolution of 12-bit or equivalent 0.97 mg with 3-axis of measurement and ∮2g of amplitude. The AD converter has also digital filters with user-defined cutoff frequencies [18]. 
The BS receives the data packets from the sensors through wireless communication, and relays the data to the PC over a USB cable. 
The PC commands and sets parameters for the network through BS. %, generate damage detection information and analyze %sensor working status such as
 %faulty sensor detection using a Java application.
 Java and Matlab are used to calculate and visualize the whole structural health condition.
%
%\subsubsection{Wireless Sensor Platform} 
	%We design and developed two additional hardware modules in our lab: 1) a radio-triggered wakeup together with synchronization module, 2) a sensor board module. The Imote2 and our designed module can be seen in Fig. Important Imote2 main board features can be observed in Table 1.  The general sensor board has digital accelerometer with 4 analog input applications. 
	%
%\subsubsection{WSN Deployment on the Designed Infrastructure}
%A twelve-story shear frame structure, made from steel, is employed as shown in Fig. 7b. 
%%The lateral stiffness of each floor originates from the four vertical steel columns, 3.81cm by 3.81cm. 
%We instrumented the structure by placing the wireless the SHM motes as shown in Fig. 6a. The 10 motes form a WSN and verify our algorithms for several rounds in several days. The test building has 10floors, at each floor; each mote is deployed to monitor the structureˇs horizontal accelerations. 
%
	In the experiment,  $R_{min}$ is adjusted by the diameter of the structure, which is adjusted by estimating the height of the test structure and each floor.
	Imote2's discrete levels of range are set to use $R_{min}$ and $R_{max}$. %We chose this so that at maximum power, a mote can talk to its up to three-hop neighbor directly. 
	%Under this range, the topology of the network along with the simplified structure is illustrated in Fig. 6c. 

\subsection{Sensor Identified Natural Frequencies}

\renewcommand{\thefigure}{G1}
\begin{figure}[t]
\begin{center}
\includegraphics[scale=0.47]{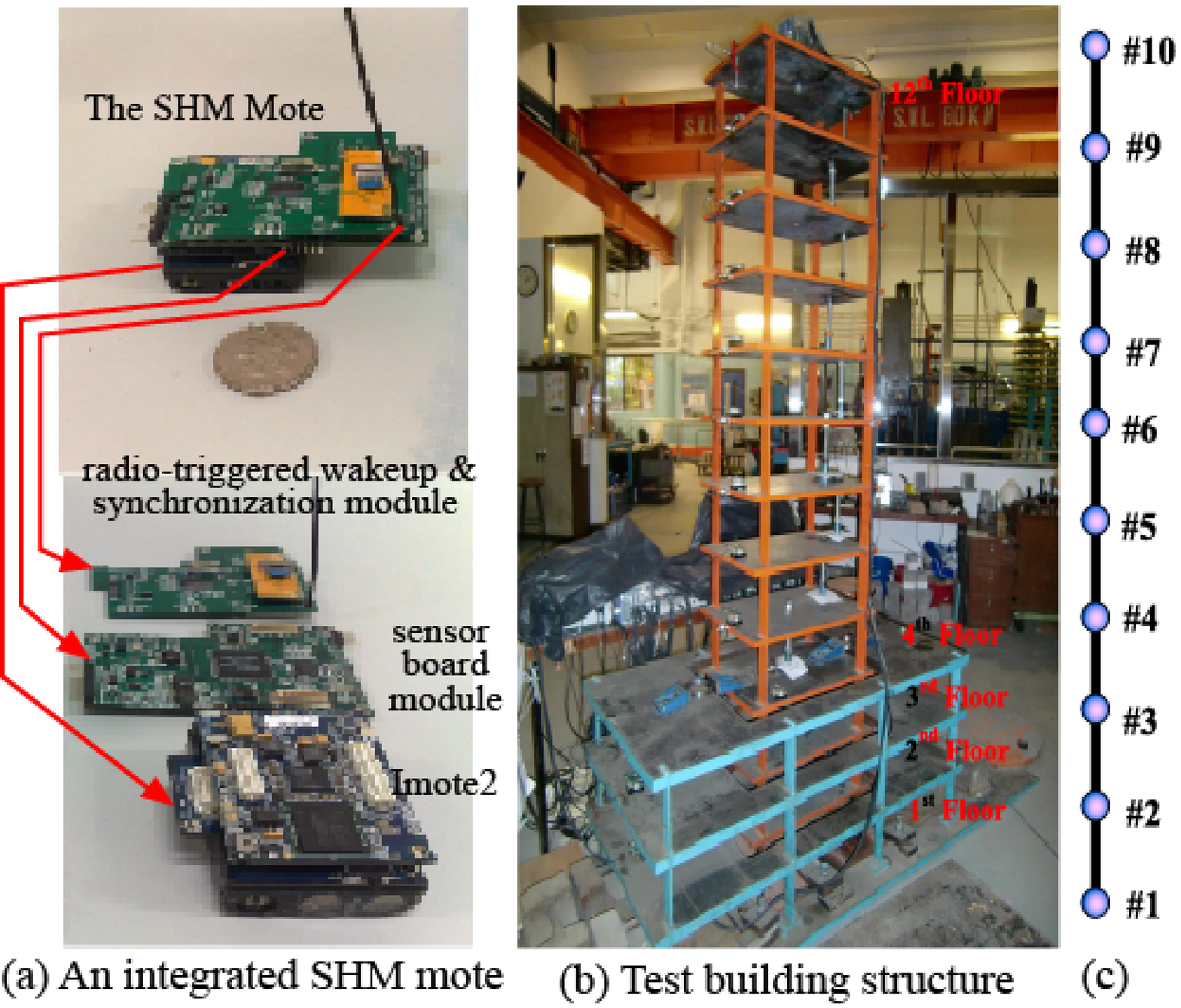}
\end{center}
\vspace {-2mm}
%\vspace {5.8cm}
\caption{ (a) The SHM mote integrated by Imote2; (b)  twelve-story test structure and the placement of 10 SHM motes on it; (c) their deployment.} 
\vspace {-2mm}
\end{figure}

\renewcommand{\thetable}{G1}
\begin{table*}[ht]
    \centering
\caption{Identified natural frequencies by the first five sensors of the experimental WSN in SPEM and DependSHM.}
    \begin{tabular}{c}
\includegraphics[scale=1.5]{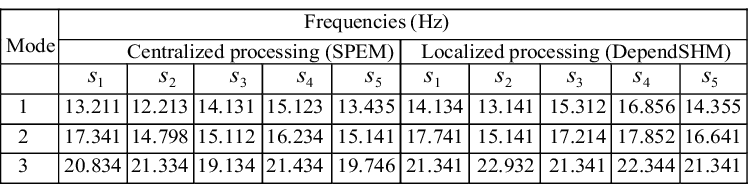}
   \end{tabular}
			 
\end{table*}

In Section 7.2.2, we have given experimental mode shapes, estimated based on natural frequencies. In the first set of experiments, we compute the natural frequencies, as shown in TABLE G1. These frequencies are used in creating mode  shapes ($\Phi$)  in the base-line structural system, when there are no damage events and no sensor faults. Note that  such a base-line mode shape should be not fixed but should be dynamic, i.e., a WSN-based SHM system can be enabled to adapt or update its base-line mode shape, taking into account dynamic environments and environmental noise factors. We find that the MII in different frequencies identified at different sensors is low (the result has been shown in Fig. 11).

%%%%%%%%%%%%%%%%%subsection%%%%%%%%%%%%%%%%%%
\subsection{Signal Reconstruction at a Faulty Sensor}

\renewcommand{\thefigure}{G2}
\begin{figure}[tb]
\begin{center}
\includegraphics[scale=1.4]{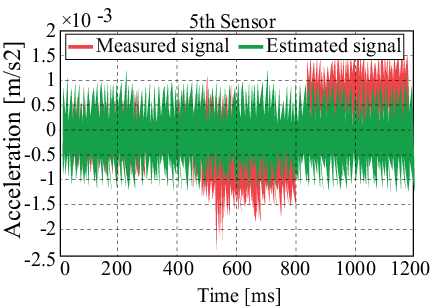}
\end{center}
%\vspace {-2mm}
\caption {Signal reconstruction of the 5th sensor (that is detected faulty).} %\label{fig2}
%\vspace {-2mm}
%\vspace {5cm}
\end{figure}

In Section 7.2.2, we have also provided the sensor fault detection results, where we have found that sensor 5 is faulty. In order to the support the results, we hereby can observe the faulty signal reconstruction of the 5th sensor, as shown in Fig. G2. The drift in the measured signal (red line) is corrected by the estimated signal (green lines). We observe the mutual independence under the fault injection at the 5th sensor. 
%Furthermore, as shown in Fig. 5, SPEM achieves some amount of MII at almost  all of the sensor locations, even when the sensors are not faulty; the amount of MII is the largest in many sensor locations in the NFMC method. For example, sensors from 1st to 3rd and 6th and 9th show some MII. It may be considered as an ``abnormal signal". It may let neighboring sensor nodes fail in detecting fault and damage. 
In \texttt{DependSHM}, when sensor nodes process data locally, the small value in the MII is achieved, ranging from 2$\%$ to 4$\%$, and they are not considered faulty. The MII provides the best value, when there is a remarkable change in  the sensor measured signals, i.e., the 5th sensor and 10th sensor are faulty. This reveals that there can a better accuracy of fault detection in \texttt{DependSHM} in practice, compared to others.

\subsection{Energy Cost $(cost(e_i))$  }
Due to space limitation, we have not presented the performance of energy cost of the WSN in Section 7.2.2, which we present in this Appendix. 

\renewcommand{\thefigure}{G3}
\begin{figure}[tb]
\begin{center}
\includegraphics[scale=1.35]{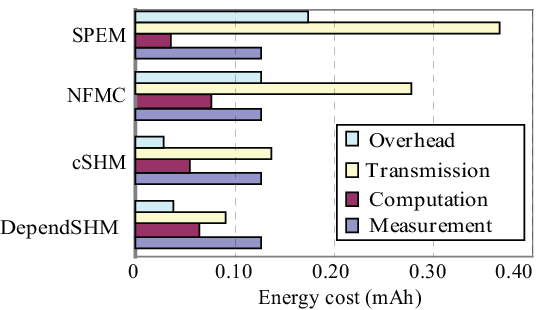}
\end{center}
\vspace {-2mm}
\caption {The performance on the energy cost of the WSN in different schemes.} %\label{fig2}
\vspace {-2mm}
%\vspace {4cm}
\end{figure}

We allow all of the sensors to sleep after each monitoring period to perform power management. The TinyOS 2.0 drivers for the Imote2 supports putting all of the hardware to sleep when it is switched off. This is obvious for a WSN-based SHM system, since a WSN does not always need to run actively  in case of specific structural event monitoring. For example, in case of aerospace vehicle monitoring, when it is not flying, the WSN may not need monitoring operations. In another case, the WSN can be scheduled to run periodically or a part of the sensors can be scheduled to wake up periodically and check health event status.  %Some common factors, such as energy cost for idle state, sleeping state are not considered in our evaluation, since they can be similar to the other systems.
 $cost(e_i)$ is calculated by the energy cost for computation, transmission, measurement, and overhead, where the overhead statistics with current cost data  %for the radio, sensor, and CPU taken from the corresponding data sheets
is combined. The data sheet can be found in \cite{1515}.

Fig. G3 shows the energy cost of a round of monitoring, $T_{d=1}$. The \texttt{DependSHM} method significantly decreases the energy cost compared to SPEM, from 0.197 mAh to 0.072 mAh. The reason is that the major energy is consumed by the raw signal transmissions to the BS. 
The actual computation cost in \texttt{DependSHM} is 0.0072 mAh to execute the basic equations and fault detection and signal reconstitution. However, it fully depends on the number of cycles that a sensor CPU requires. It also varies from sensor to sensor based on the tasks a needs to do. A sensor does not need computation for signal reconstruction if there is not fault. In such a case, a sensor can save an average of 0.0027 mAh. More importantly, in \texttt{DependSys}, the computation saves the Imote2 an average of 0.165 mAh during transmission, since it reduces the time that the CC2420 radio is active%and transmitting by 9369ms
. 
The overhead is caused by end-to-end transmission delay and writing/reading data to/from Imote2's memory, since we depend on local processing. In both SPEM and NFMC methods, transmitting a large amount of raw data in each $T_d$ (i.e., transmission of natural frequency sets and frequent retransmissions caused by packet losses) increases $cost(e_i)$. However, NFMC achieves slightly lower energy cost for transmission than SPEM.

\renewcommand{\thefigure}{G4}
\begin{figure}[tb]
\begin{center}
\includegraphics[scale=1.4]{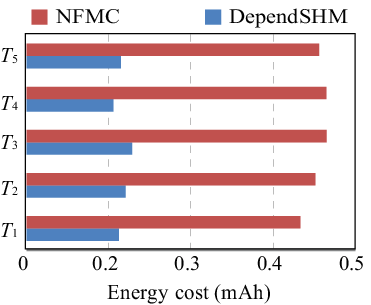}
\end{center}
\vspace {-2mm}
\caption {The performance on the energy cost of the WSN in the first five round of monitoring in \texttt{DependSHM} and NFMC.} %\label{fig2}
\vspace {-2mm}
%\vspace {4cm}
\end{figure}

 Further performance analysis of $cost(e_i)$ in five rounds of monitoring ($T_d, d=1,...,5$) can be seen in Fig. G4. This shows the actual amount of energy cost required in  \texttt{DependSHM}. We can see that  \texttt{DependSHM}  outperforms NFMC significantly because of the above causes, cluster maintenance,  and network maintenance (e.g., faulty sensor isolation), particularly the set of  mode shapes transmitted from the cluster-head to the BS. % still incur energy cost. 
This is because the final  mode shapes of each cluster is transmitted by each cluster-head, while SPEM requires transmission of all natural frequency sets. 
 In our distributed solution, there is no frequent retransmission and the final mode shapes transmitted by each sensor are without sensor fault information. 
% , since the BS can receive the final results through a neighbor of a sensor. 
%In our scheme, the overhead is also caused by the data writing to and copying from the memory. 
%The total energy cost in each period of monitoring can be seen in Fig. 11(b). %, where $T =5t, i.e., five rounds of monitoring.
 %After a round of monitoring is over, the network goes to sleep until next monitoring period. 
% 
In the case of faulty sensor detection and signal reconstruction, the system consumes a small amount of energy in computation with a slight overhead, which is 5$\%$ to 8$\%$ of the total energy cost in each round. 

%The result is shown in Fig . ?? 
In a concluding remark about the results we have found and presented in this paper, our proposed dependable, distributed SHM solution outperforms centralized solution %(most of the existing solution such as [][] are centralized) 
almost in all aspects, including, energy cost of the WSN and offering monitoring system dependability.

\end{document}